\documentstyle[12pt,aaspp4,psfig]{article}

\newcommand{\zsol}{$Z_{\odot}$}
\newcommand{\NH}{$N_{\rm H}$}

\newcommand{\asca}{{\it ASCA}}
\newcommand{\rosat}{{\it ROSAT}}
\newcommand{\nix}{$\cdot\cdot\cdot$}

\lefthead{T. Heckman et al.}
\righthead{Observations of Arp 299}

\slugcomment{.}

\begin{document}

\title{An X-ray and Optical Investigation of the Starburst-driven Superwind in
the Galaxy Merger Arp 299}

\author{T. M. Heckman$^1$}
\affil{Department of Physics and Astronomy, Johns Hopkins
University, Homewood Campus, 3400 North Charles Street, Baltimore, Maryland 21218}
\author{L. Armus$^1$}
\affil{SIRTF Science Center, California Institute of Technology, 100-22, Pasadena, California 91125} 
\and
\author{K. A. Weaver and J. Wang} 
\affil{Department of Physics and Astronomy, Johns Hopkins
University, Homewood Campus, 3400 North Charles Street, Baltimore, Maryland 21218}

\parindent=0em
\vspace{5cm}

1. Visiting astronomers, Kitt Peak National Observatory, NOAO,
operated by AURA, Inc. under cooperative agreement with the
National Science Foundation.
\newpage

\parindent=2em

\begin{abstract}
We present a detailed investigation of the X-ray and optical 
properties of the starburst-merger system Arp299 (NGC 3690, Mrk 171), with
an emphasis on
its spectacular
gaseous nebula. We analyse \rosat\ and \asca\ X-ray data and
optical spectra and narrow-band images.
The X-ray nebula in Arp 299  has a diameter of $\approx$45
kpc, is elongated roughly along the HI minor axis of the merging
system, has
an outer (inner) temperature of 2.3 (9) million K, a luminosity of
$\sim2\times10^{41}$ erg s$^{-1}$, a gas mass of 7$\times10^9$ M$_{\odot}$,
and a thermal energy content of $\sim$10$^{58}$ ergs.
An additional hard X-ray component
is present with a luminosity in the 0.1 to 10 keV band of $\sim4
\times10^{41}$ erg s$^{-1}$ most likely due to X-ray binaries or possibly
inverse Compton scattering. The overall X-ray spectrum is consistent with that
of other starbursts (Dahlem, Weaver \& Heckman). Compared to the X-ray nebula,
the optical emission-line nebula is smaller (20
kpc) but much more luminous ($\sim 10^{44}$ erg s$^{-1}$ $\approx$ 4\%
 L$_{bol}$ for Arp 299).
The kinematics of the
emission-line nebula are complex, with the line-widths increasing
systematically with decreasing H$\alpha$ surface-brightness to values
of 500 to 700
km s$^{-1}$ in the faint outer filaments. The relative strengths of the
low-ionization forbidden lines ([SII],[NII], and [OI]) also increase
systematically as the gas surface-brightness decreases and as the
line widths increase.
We measure high gas
pressures in the inner emission-line nebula (P/k $\sim$ few $\times$ 10$^6$
K cm$^{-3}$) that
decline systematically
with increasing
radius.

We suggest that the on-going galaxy collision has tidally-redistributed the
ISM of the merging galaxies. The optical emission-line nebula results as this
gas is photoionized by radiation
that escapes from the
starburst, and is shock-heated, accelerated, and pressurized by a
`superwind' driven
by the collective effect of the starburst supernovae and stellar
winds. Since empirically only those galaxy mergers that
that contain luminous starbursts have
bright X-ray nebulae, this implies that the luminous X-ray nebula
in Arp 299 is also powered by the starburst outflow (rather than by the
collisions of gas clouds during the merger).
This outflow can be most directly traced by its X-ray emission, which
is plausibly a mass-loaded flow ($\sim 10^2$ M$_{\odot}$ yr$^{-1}$) of
adiabatically-cooling gas that
carries out a substantial fraction of the energy and metals injected
by the starburst at a speed close to the escape velocity from
Arp 299.  The mass outflow rate exceeds the star-formation rate
in this system. We conclude
that powerful starbursts are able to heat (and possibly eject) a significant
fraction of the interstellar medium in galaxies that are the products
of mergers.

\end{abstract}
 

\newpage
 

\section{Introduction}

\subsection{General Motivation}

Galaxies that emit the bulk of their luminosity in the far-infrared
(hereafter far-IR galaxies) are an energetically significant
component of the local universe (e.g Soifer et al 1987;
Sanders \& Mirabel 1997). They are also the only local systems
(apart from QSOs) that
have bolometric luminosities similar to typical high-redshift galaxies
selected by UV (Steidel et al 1996; Meurer, Heckman, \& Calzetti 1998)
or sub-millimeter
(e.g. Hughes et al 1998; Blain et al 1998)
techniques. As such, they offer unique laboratories to study the
physical processes that were important in the formation and early
evolution of galaxies (Heckman 1998).

One of
their most interesting attributes are their galactic-scale and
highly luminous gaseous emission-line nebulae (cf. Armus, Heckman,
\& Miley 1990; Veilleux et al.\ 1995). In several
well-known cases like M 82 and NGC 253, hot X-ray emitting gas has
also been discovered, and this gas shows a clear morphological
relationship to the optical emission-line nebula (cf. Dahlem, Weaver,
\& Heckman 1998 and references therein).

It seems clear in these
cases that the X-ray nebulae are energized by the mechanical energy of
an outflow driven by the multiple supernovae in the starburst (a
`superwind' --- cf. Heckman, Lehnert, \& Armus 1993; Bland-Hawthorn 1995).
It would be important
to establish the ubiquity of the superwind phenomenon in far-IR
galaxies, since such outflows could play an important role in
heating and chemically-enriching galactic halos and the inter-galactic
medium (cf. Heckman, Armus, \& Miley 1990; Gibson, Loewensten, \& Mushotzky
1997).

The X-ray nebulae studied in detail and at high spatial resolution 
to date have been mostly
associated with far-IR galaxies of modest luminosity ($L <$ few $\times
10^{10} L_{\sun}$). The situation may be different and more complex for
more powerful far-IR galaxies. These systems are
also primarily powered by
intense starbursts (Genzel et al 1998) although deeply-buried AGN
are energetically significant in some cases and may dominate the
systems with the highest bolometric luminosities (cf. Sanders \& Mirabel
1997 and references therein). They
are created in the aftermath of a galactic collision or merger (cf.
Mihos \& Hernquist 1994), so that
the dissipation of the kinetic energy
of the colliding interstellar media may be energetically significant (Harwit
et al.\ 1987). It is also particularly interesting to study the effects
of possible superwinds on the interstellar media of merging
disk galaxies in general, since such hydrodynamical processes may be important
in the subsequent transformation of the merger product into an
elliptical galaxy (e.g. Schweizer 1992).

We have therefore undertaken an analysis of X-ray and optical data
for a small sample of some of the nearest and brightest powerful infrared
galaxies (L$_{FIR}$ $\geq$ 10$^{45}$ erg s$^{-1}$)
in order to determine the characteristics of their
nebulae and thereby determine the nature of their energy sources.
Heckman et al.\ (1996) presented X-ray and optical data on the
prototypical `ultraluminous' far-IR galaxy Arp 220, and showed that
this galaxy appears to have an outflow that is qualitatively
similar to (but more powerful than) the outflows seen in M 82 and
NGC 253. In Wang  et al.\ (1997), we reported the results for
the case of Mrk 266, and here we discuss a
third example. Iwasawa \&
Comastri (1998) have recently reported on the X-ray properties of the
IR-bright merger system NGC 6240.

Our approach is complementary to the recent work by Read \& Ponman (1998),
who have systematically studied the X-ray properties of an optically-
selected sample of merging galaxies using \rosat\ data.
The members of their sample
are not in general
powerful starbursts, and only Arp 220 meets the above far-IR
luminosity criterion.

\subsection{Arp 299}

The powerful far-IR system Arp 299 (also known as NGC 3690 and Mrk 171)
consists of two overlapping galaxies
that appear to be in the late stages of a merger. 
\footnote{In most past papers on this system, the Eastern member of the
merging galaxy pair
has been designated IC 694 and the Western member as NGC 3690. However,
as pointed out in IAU Circular 6859, IC 694 is properly the designation
of a small E/S0 galaxy located about 1$'$ to the northwest
of the merging galaxy pair, while NGC 3690 properly refers to the
merging pair. To avoid further confusion, we will adopt the name Arp 299
to refer to the system rather than NGC 3690.}
\footnote{In this paper, we will adopt a distance to Arp 299 of 44 Mpc,
derived using the linear Virgocentric infall model described in
Marlowe et al.\ (1995) and a far-field value of H$_{0}$ = 75 km s$^{-1}$
Mpc$^{-1}$. This implies a scale of 1\arcsec = 0.21 kpc and 1\arcmin = 13 kpc.}
It has one of the brightest, most luminous, and largest emission-line nebulae
in the
extensive H$\alpha$ imaging survey of far-IR galaxies by AHM90. As we
will describe in this paper, this material is immersed in and
surrounded by a large and bright region of soft X-ray emission. Arp 299
is therefore a prime target for a detailed examination of the 
relationship between the warm and hot gaseous components in powerful
far infrared galaxies.

Arp 299 has been extensively studied following the landmark
paper
of Gerhz, Sramek, \& Weedman (1983): in the X-ray (Zezas, Georgantopoulos,
\& Ward 1998 - hereafter ZGW), 
UV (Meurer et al 1995), optical (Friedman et al 1987),
IR (Fischer et al 1983; Telesco, Decher, \& Gatley 1985;
Beck et al 1986; Nakagawa et al 1989; Joy et al 1989;
Wynn-Williams et al 1991),
and radio (Condon 1991;
Lonsdale, Lonsdale, \& Smith 1992; Smith, Lonsdale, \& Lonsdale
1998) 
wavebands, and in HI (Stanford \& Wood 1989
; Hibbard 1997), OH
(Baan \& Haschick 1990), CO (Aalto et al 1997; Sargent \& Scoville 1991),
and HCN (Aalto et al 1997). It has been the site of several recent
supernovae (SN1992bu - IAU Circular 5960; SN1993G - IAU Circular 5718;
SN1998T - IAU Circular 6859; and a possible radio supernova described
in IAU Circular 4988).
Arp 299 has a
bolometric luminosity of approximately 6 $\times$ 10$^{11}$ L$_{\sun}$,
which is dominated by far-IR emission. $ISO$ spectroscopy (Genzel
et al. 1998) shows that even in the mid/far-IR there is no sign of
a substantial contribution to the energetics of Arp 299 by a dust-buried
AGN. As we will report below, this is also true in the hard X-ray band
probed by $ASCA$. Thus, Arp 299's impressive luminosity seems to be
provided largely or entirely by stars.

The centers of the two merging galaxies in Arp 299 are separated by only about
22\arcsec (4.5 kpc in projection), with a clear overlap in the disks.
Following
the nomenclature of Gerhz, Sramek, \& Weedman (1983), the compact site
of activity in the central region of the Eastern galaxy is denoted `A', while
the activity in the Western galaxy breaks up into a complex of sources (a double
source ``B1", and ``B2" the first of which is evidently associated with the nucleus of
the galaxy, and a third source ``C" roughly 8\arcsec or 1.6 kpc NNW of B1).

In the present paper we present an analysis of both X-ray and
optical imaging and spectroscopy of Arp 299. Our approach complements that
of ZGW, who analyzed the PSPC spectrum, but not the morphological
structure of the \rosat\ data.
Our data
analysis is described in \S 2 and the results for X-ray and optical
data are presented in \S 3 and \S 4 respectively. We will
discuss and interpret these results in \S 5 in an attempt to
determine how the bright optical/X-ray nebula is heated and how
Arp 299 and its nebula may fit in to the general class of powerful
far-IR galaxies. In \S 6 we will summarize our conclusions.

\section{Observations and Data Reduction}\label{sec:observations and Data Reduction}

\subsection{X-ray Imaging and Spectroscopy} 

Arp 299 has been observed in the soft and medium-energy X-ray bands 
with the Rontgensatellit (\rosat) and the Advanced Satellite for
Cosmology and Astrophysics (\asca; Tanaka, Inoue \& Holt 1994).
\rosat\ contains the high-resolution soft X-ray imaging 
detector (HRI) and the positional sensitive
proportional counter (PSPC), while \asca\ contains 
two sets of detectors which are the solid-state imaging
spectrometers (SIS; designated here as S0 and S1) and the gas imaging
spectrometers (GIS; designated S2 and S3). 
The data from the two satellites are complementary.
For soft X-rays ($0.1-2.0$ keV), \rosat\ provides $\sim 5''$
spatial resolution with the HRI and moderate
spectral resolution with the PSPC.  The \asca\ instruments 
extend the spectral coverage to 10 keV with a factor 
of ten better spectral resolution (albeit with poor 
spatial resolution).  The X-ray data were obtained by 
us (\rosat) and from the HEASARC\footnote{The high energy 
astrophysics data archive at http://heasarc.gsfc.nasa.gov/} operated 
by the LHEA\footnote{The Laboratory for High Energy
Astrophysics.} at the Goddard Space Flight Center (\asca).

The \rosat\ HRI observations occured on 
April $13-14$ 1992 for a total live time integration of 6691
seconds.  The nominal pointing center of the HRI was
11h:25m:42.8s, $+$58d:50m:15s, which places the center of 
the HRI field of view at a position approximately halfway
between the interacting galaxies that comprise Arp 299.
The HRI background is approximately constant over the field of view
and so we subtract a constant offset, which is determined 
by averaging the background at different source-free locations. 
After background subtraction, there are 285 counts 
in the HRI image, corresponding to a count rate of 
0.0426 cnt s$^{-1}$ from Arp 299.  For imaging analysis, 
the data were binned and smoothed with a Gaussian having 
$\sigma=3''$, yielding an effective resolution of $7''$ FWHM.

Arp 299 was observed with the PSPC in multiple intervals from 1991
November 19 to 1993 April 23 for 9,925 net seconds.  Times of
a high master veto rate (MVR above $\sim$220) indicating charged
particle contamination were negligible, and so we include all 
of the on-source data in our analysis.  After correcting for 
vignetting and subtracting background, the PSPC count 
rate for Arp 299 is
0.10 counts s$^{-1}$, yielding 983 counts.    

\asca\ observed Arp 299 on April 16 and December 1 1994
for $\sim6$ and $\sim36$ ks, respectively.  The first 
data set provides few source counts and so we examine 
only the data from the second observation.  
To remove times of high background,
we selected data within time ranges defined by the 
following criteria: for the SIS, the satellite was
required to be outside the SAA with an Earth elevation 
angle greater than 10$^{\circ}$ and a geomagnetic cutoff rigidity 
(COR) greater than 6 Gev c$^{-1}$.  In addition, we removed 
times less than 16 s after an SAA passage, 60 s after a 
day-to-night passage, and during times of Bright Earth when 
the elevation angle was less than 20$^{\circ}$.  For the 
GIS, the satellite was
required to be outside the SAA with an Earth elevation 
angle greater than 5$^{\circ}$ and a geomagnetic cutoff rigidity 
(COR) greater than 6 Gev c$^{-1}$.  Two background 
spikes were also removed.  Remaining on-source 
times were 26.3 ks (S0), 28.3 ks (S1) and
36.2 ks (G2 and G3).   The background during the observation   
ranged from 20 to 30\% of the total count rate 
from $1-10$ keV.  Source spectra were extracted  
by selecting circular regions of $\sim3'$ radius and 
background was selected from  
near the source in the GIS and from the same chip 
in the SIS.  The background-subtracted count rates 
are 0.042, 0.039, 0.023, and 0.033 counts s$^{-1}$,
for S0, S1, S2, and S3, respectively
(totalling 4,234 source counts).

To extract \rosat\
images and spectra we use the IRAF/PROS V2.3.1 data reduction package.
For \asca\, we use FTOOLS v4.0. 
With the exception of the spatially-resolved halo
emission in the PSPC, all spectra
are grouped to have $\ge15$ counts per bin to allow the use
of $\chi^2$ statistics.   There are too few counts in the   
halo to assume a Gaussian probability distribution  
and so Poisson errors and the C-statistic (Cash
1979) are used for spectral modeling.

\subsection{Optical Spectroscopy}

Optical spectra of the Arp 299 system were obtained in May 1987 and
January 1988 using the KPNO 4-meter Mayall telescope in combination with
the RC spectrograph and a TI2 CCD.  The May 1987 data (hereafter May87)
are of medium spectral resolution (3.6$\AA$ Full Width Half Maximum - FWHM)
and cover the wavelength range of 6000-7000 $\AA$.  Three slits (2.0$''
\times 3.0'$) were placed at a position angle (PA) of 90$^{o}$ along the
Arp 299 system.  Two of these slits intercept components ``C" and ``B"1
respectively (in the notation of Gehrz, Sramek, \& Weedman 1983).
The third slit was placed to be parallel to the first two,
but displaced 24$''$ north of component "C''.  The January 1988 data
(hereafter Jan88) are of low resolution (13.5$\AA$ FWHM) and
cover the wavelength range of 4200-7000$\AA$.  The slit dimensions were
the same as those used for the May87 data., but here the slits were placed to
intercept component "A'' (at a PA=90$^{o}$) and to pass along the tail of
emission to the northwest of Arp 299 (at a PA=120$^{o}$).  Both the May87
and the Jan88 data were binned by a factor of 2.0 in the spatial direction
to achieve a pixel scale of 0.89$''$ pixel $^{-1}$.  

The May87 and Jan88 spectra were bias-subtracted and flat-fielded using
the NOAO's IRAF software. The LONGSLIT package was used for wavelength
calibration and distortion correction of the raw spectra.  Only the low
resolution spectra (Jan88)
were taken under photometric conditions, and these were
flux-calibrated in IRAF using KPNO IIDS standard stars.  
For each extracted spectrum, the SPLOT package was used to measure the
integrated line fluxes, central wavelengths, linewidths, equivalent widths
and continuum levels.  The H$\alpha+$[NII] complex was deblended in the
data by requiring the widths of the lines to be the same.  The [SII] lines
at 6717\AA ~and 6731\AA ~were required to have the same linewidth and, in
regions of low signal to noise, to have a fixed wavelength separation of
14.6\AA.  In all cases, the uncertainties in the line fits were estimated as
a function of signal-to-noise ratio by comparing the measured value of the
[NII]$\lambda\lambda$ 6584\AA/6548\AA ~flux ratio to that of the value of
3.0 expected
from atomic physics.  


\section{Results}\label{sec:results}

\subsection{X-Ray Morphology}

\subsubsection{\rosat\ PSPC}

The background-subtracted $0.1-2.4$ keV  
PSPC image of Arp 299 is shown
superimposed on the Palomar Sky Survey image in Figure 1.  Accurate
astrometry is not possible for the PSPC because no independent X-ray
point sources are present in the field of view, and so the image is 
aligned with the optical centroid of the Arp 299 system to within the
accuracy of the \rosat\ pointing uncertainties ($\sim4-6''$).

In order to investigate the morphology of the X-ray image as a function of
energy, we constructed ``soft'' band ($0.1-0.5$ keV) and ``hard'' 
band ($0.5-2.4$ keV) images.  Photons were selected
from the appropriate pulse invariant (PI) channels
and the central $6.17'$ by $6.17'$ regions were extracted from the original
unbinned X-ray map ($0.5''$ pixels).  These images were then
smoothed with a Gaussian function with $\sigma = 30$ pixels
($35''$ FWHM).  After smoothing, the images
were shifted slightly to align the center of the hard \rosat\ 
image (which represents the core of X-ray emission from 
the galaxy) with the
center of emission mapped with the HRI (see below).

The PSPC images, shown in Figure 2, reveal that the hard and soft X-ray
isophotes are elongated in nearly orthogonal directions.  The hard X-ray
image is more compact, possibly oriented in the east-west direction (no doubt
resulting from the complex of point sources seen in the HRI data).  The soft
X-ray isophotes on the other hand, are clearly oriented in the 
NNE/SSW direction, and are 
extended on a scale of $\sim 3.5'$, or 45 kpc. The size of the soft
X-ray emission is similar to that of the HI mapped by Stanford \& Wood
(1989), but the X-ray emission is roughly perpendicular
to the NW/SE orientation of the HI. The shapes of
the soft-band and hard-band images imply at least two distinct emission regions.

\subsubsection{HRI}

The HRI image (Fig. 3) has a total background-subtracted count rate of 0.0426
cnts s$^{-1}$, and shows a complex, clearly resolved structure
extended over more than an arcminute in a predominantly east-west
direction.  This is the direction of the two galaxies in the interacting
system.  There are numerous
local maxima in the X-ray image, but the brightest source to the east is only
$\sim10''$ north of the non-thermal radio and near-infrared nucleus of
the Eastern galaxy
(Gehrz, Sramek \& Weedman 1983; Condon 1991; Wynn-Williams, et al.
1991).  This is nucleus ``A'' in the nomenclature of Gehrz, Sramek \&
Weedman (1983).  Given the pointing accuracy of the \rosat\, and the fact
that nucleus A is the single primary peak in both the radio and near
infrared data, we hereafter identify the eastern X-ray peak with nucleus A.
The western X-ray peak is extended in a ``ridge'' of emission
with two prominent knots separated by 4 to 5 $''$.
When the eastern peak is aligned with nucleus ``A", the southeast knot in the
ridge is coincident with component ``B1" and the northwest knot
appears to be associated with a bright knot of H$\alpha$ emission a few
$''$ to the west of component ``B2".
Note that while B1 is the peak in the 2.2 micron, 10 micron, and radio
emission in the Western galaxy, ``B2" is the optical continuum peak.
Finally, the infrared and radio peak
known as component ``C" 
is not located at the position of a relative peak in the
\rosat\ HRI image (a weak X-ray source tabulated by ZGW is located about
5$''$ to the NW of C). Thus,
although
component ``C" is a source of optical, near-infrared and radio emission, it
does not appear to be a strong source of X-rays.


We can estimate the contribution of point sources to the total number of
counts in the HRI image by summing the flux from the three
strongest point sources within 5$''$ square apertures
(a ``detection cell'' that includes 50\% of the total flux expected from a point
source). 
These three X-ray sources are equally bright (to within $\pm$10\%) and
together account for approximately 21\%
of the total counts in the HRI image, after correction for the 50\% of the
flux beyond the 5$''$ detection cells (0.0087 counts s$^{-1}$).
Although the HRI data do not provide much spectral information,
we can estimate their fluxes and luminosities over $0.1-2.4$ keV range
by assuming a simple spectral model (David et al 1997).
If the point sources
are modelled as a MEKAL plasma with kT $=$ 0.8 keV 
(see \S 3.2.2) and line of
sight neutral HI column of N$_H=10^{20}$ cm$^{-2}$ (N$_H=5\times10^{21}$
cm$^{-2}$), they have a combined unabsorbed X-ray luminosity of about
$0.6\times10^{41}$ erg s$^{-1}$ ($2.2\times10^{41}$ erg s$^{-1}$) from
$0.1-2.4$ keV.  Alternatively, if the point sources are modelled as having
power law spectral shapes with a photon index of 1.6 
(the index of the hard X-ray component - see \S 3.2.3) they have a combined
X-ray luminosity of $1.0\times10^{41}$ erg s$^{-1}$ (3.1$\times10^{41}$
erg s$^{-1}$) from $0.1-2.4$ keV.

Each of the three sources thus has a luminosity in the 0.1 to 2.4 keV band of
$\sim$ few $\times 10^{40}$ to 10$^{41}$ erg s$^{-1}$. This can be 
compared to the most powerful individual X-ray sources known in nearby
galaxies, which have peak
X-ray luminosities of $\sim$ 10$^{40}$ erg s$^{-1}$: SN1978C in NGC 1313
(Ryder et al 1993), SN1986J in NGC 891
(Bregman \& Pildis
1992), SN 1993J in M 81 (Zimmermann et al 1994; Kohmura et al 1994), and the
highly variable sources in the nuclei of M 82 (Collura et al
1994) and NGC 3628 (Dahlem, Heckman, \& Fabbiano 1995). It is worth emphasizing
that at the distance of Arp 299, 5$''$ corresponds to roughly 1 kpc, and
that the sources are coincident with the sites of the most intense
star-formation in Arp 299. Higher resolution X-ray imaging with AXAF may well
resolve these sources into multiple sub-components.

\subsection{X-ray Spectra}

The different morphologies of the hard and soft-band 
PSPC images suggest that there are 
at least two sources of X-rays in the Arp 299 system with
different spectral shapes and possibly different emission
mechanisms.  To uncover these emission mechanisms we extract  
spatially-resolved spectra from the central and outer regions.  The  
central or ``core'' region is defined with  
a circle of diameter = 1.3$'$ centered on the X-ray peak 
of Fig. 2a.  The outer or ``halo'' region is defined with an ellipse 
having a major axis = 2.78$'$, minor axis = 2.22$'$, and P.A.
$=193^{\circ}$ centered on the X-ray peak of Figure 2a.
Background is taken from source-free regions
near the galaxy.  The extraction regions, source counts 
and count rates are listed in Table 1 and the 
spatially-resolved halo and core spectra are 
shown in Figure 4.  

\subsubsection {The Halo Spectrum}
 
There are 328 counts from within the halo region, 119 of which are source 
counts and 209 (almost 2/3) of which are background counts. 
We estimate the contribution of  
background counts to the total spectrum by calculating the 
areas of the halo and background regions and then 
using the ratio of areas to predict the 
contribution of background in the halo region. 
We then determine an empirical spectral model for the background
and include this model (properly normalized) in our spectral 
fits to ``subtract'' the background contribution. 
Our empirical background model consists of a power law
with $\Gamma$ = 2.3 $\pm$0.2 and an unresolved Gaussian-shaped emission 
line at 0.53 keV.  The line represents oxygen fluorescence from 
Earth's atmosphere and the power law 
approximates contributions from the particle background and 
diffuse X-ray background, as well as long-term and short-term 
enhancements in the background rate. 
  
The statistics are too poor to distinguish between different physical
models for the halo emission.  In particular, the data cannot distinguish
between the two most natural choices, which are 
a power-law spectrum (where the X-rays are
produced by inverse-Compton scattering of lower energy photons) and a
thermal spectrum (where the X-rays are produced in a plasma that has been
heated by shocks). A
power-law spectrum is expected 
if the soft X-rays result from emission associated
with an extended synchrotron-emitting radio halo, such as that found
around M82 (Seaquist \& Odegard 1991).  However, the radio 
emission in Arp 299 is not extended on the same scale as the 
soft X-rays (Condon et al. 1990).  In fact, the soft X-ray emission is
extended in a direction (North-South) that is roughly perdendicular
to the radio source major axis.
This implies that the X-ray emission is not related to the 
emission at radio wavelengths in Arp 299,
and hence a power-law description of the halo is unlikely.
Recent studies of extended emission in starburst galaxies
indicate that the soft emission is primarily thermal
(see Dahlem, Weaver \& Heckman 1998) and so we
choose a model that consists of an emission 
spectrum from hot, diffuse gas based on  
calculations by Mewe and Kaastra with 
Fe L calculations by Liedahl (Mewe, Gronenschild \& van den Oord
1985; Mewe, Lemen \& van den Oord 1986; Kaastra 1992 - 
hereafter referred to as the MEKAL model).
  
When the data are fitted with a single temperature
MEKAL model and solar abundances are assumed, we find no 
evidence for absorption in excess of the Galactic column density
(0.9$\times$10$^{20}$ cm$^{-2}$; Burstein \& Heiles 1982), and the  
spectrum is well described with a plasma having
kT$=0.20^{+0.12}_{-0.07}$ keV (Table 2).    
The $0.1-2.4$ keV unabsorbed flux for this model
is $2.8\times10^{-13}$ ergs cm$^{-2}$
s$^{-1}$, corresponding to a halo luminosity of $\sim7\times10^{40}$ erg
s$^{-1}$.

\subsubsection{The Core Spectrum}

For the core spectrum, a single-component power-law  
model yields a poor fit (Table 2; Model A) and results in
positive residuals near 0.9 keV
that resemble a blend of primarily Fe L and Ne 
line emission (Figure 5a).  MEKAL plasma models 
with solar abundances also yield poor fits
(Table 2; Models B1 and B2),  even when the absorbing  
column density is a free parameter 
in the fit.  This can be seen in 
Figure 5b, which shows evidence for excess 
flux toward higher energies.  If the abundance is allowed 
to vary, the fit improves (Table 2; Model B3) implying an 
abundance of 0.12(+0.10 -0.06)\zsol. 
Such a low abundance seems highly unusual, especially in 
light of the intense starburst in Arp 299,
but since 
abundance calculations are derived from
line equivalent widths, the possibility exists that a 
second thermal or hard X-ray component (not uniquely detectable 
with the PSPC) may contribute
continuum flux at low energies.  Such a component would 
dilute the line emission from 
the plasma and cause the abundance of the
hot gas to be underestimated.
This behavior is typical of starburst spectra that
intrinsically contain multiple emission components
but are fitted with single-component plasma models (Dahlem, Weaver, \&
Heckman 1998; Weaver, Heckman, \& Dahlem 1998). 
For the remainder of this paper we therefore assume that
the X-ray emitting gas has solar abundance.

For the case of solar abundances, the PSPC data require 
at least two spectral 
components.  Models that consists of a medium-temperature MEKAL 
plasma and a thermal bremsstrahlung with kT
= 7.0 keV (Table 2; Model C) or a medium-temperature
MEKAL plasma and a power law with $\Gamma$=1.7 (not shown) 
provide excellent fits as long as the column density is 
allowed to be about 2 to 3 times larger than 
the Galactic value.  In this case,  kT$\sim 0.7$ keV  
for the medium-temperature plasma.  
The ratio of the core spectrum to the
two-component MEKAL plus bremsstrahlung model is
shown in Figure 5c. The $0.1-2.4$ keV unabsorbed flux for this model
is $1\times10^{-12}$ ergs cm$^{-2}$
s$^{-1}$, corresponding to a core luminosity of $\sim2.3\times10^{41}$ erg
s$^{-1}$.

\subsubsection {Joint PSPC plus \asca\ fits ($0.1-10$ keV)}

\asca\ cannot spatially resolve the X-ray emission and so we use 
the integral PSPC spectrum for joint fits.  The spectral 
coverage to $\sim10$ keV provided by \asca\ allows us 
to detect a distinct hard X-ray component. 
Single power-law or MEKAL plasma models provide
extremely poor fits to the \asca\ plus PSPC spectrum (Table 3; 
Models A and B).  Note that this is true even if the abundance is 
a free parameter in the fits.  The ratio
of the joint data to the MEKAL model (Figure 6) clearly 
illustrates both excess line emission near 0.9 keV and a 
high energy excess beyond about 5 keV.
A two-component MEKAL plus power-law model provides the
best fit (Model C) but the MEKAL temperature is higher than 
that found for the halo (\S 3.2.1) and the derived abundances are 
sub-solar.  The abundances can be reconciled with solar if 
we add to the model the 0.2 keV MEKAL plasma component 
found in the halo (Model D).  In this case, the hard  
X-ray component requires absorption in excess of the Galactic value.

We note that our approach and results differ somewhat from ZGW
who also fit the joint \asca\ plus PSPC spectrum. They did not
require our soft (0.2 keV) component, nor did they find evidence for
intrinsic absorption. The inclusion of the soft component in our model
fits is dictated primarily by our spatial analysis of the PSPC
image as a function of energy (which ZGW did not undertake). Our
analysis demonstrates that the soft 0.2 keV component is definitely
present in the halo (and hence must be incorporated in the global
fit to the \asca\ plus PSPC spectrum). As described in the previous
paragraph, inclusion of the 0.2 keV component then leads to
higher values of N$_H$ for the hard component. This
nicely illustrates the power of combining all the available
information provided by the X-ray images and spectra when interpreting
the complex X-ray emission from starbursts.

The data and best-fitting three component model (Model D) are shown 
together in Figure 7 and the deconvolved model 
is shown at full resolution in Figure 8.  The goodness of 
the fit is illustrated by the ratio of the data to 
the best-fitting model (Model D) in Figure 9. In addition to
the 0.2 keV halo component, the second thermal component has
a temperature of 0.77 keV, and the hard component 
is well represented as
either a power-law with photon index 1.6 or thermal 
bremsstrahlung emission with kT $\sim$ 11 keV.

The intrinsic column of HI in Model D (of-order $10^{21}$ cm$^{-2}$) is similar
to that inferred from the dust extinction measured in our optical spectra
(see section 4.2 below).  The
unabsorbed fluxes for the $0.1-2.4$ keV and $2-10$ keV bands
are $\sim1.4\times10^{-12}$ erg cm$^{-2}$ s$^{-1}$ and
$\sim1.1\times10^{-12}$ erg cm$^{-2}$ s$^{-1}$, respectively
(Table 4).  The corresponding X-ray luminosities are
3.3$\times10^{41}$ erg s$^{-1}$ and 2.5$\times10^{41}$ erg s$^{-1}$.

\subsubsection {The Nature of the Hard Component}

We will discuss the nature of the two MEKAL plasma components extensively
in sections 3.3 and 5.2 below. Here we remark that the hard (powerlaw
or bremsstrahlung) component is most likely due to a combination
of inverse Compton scattering of soft starburst photons by the relativistic
electrons responsible for the radio synchrotron emission (see Moran \&
Lehnert 1997) and a population of X-ray binaries. Comparing the flux
of this hard component in the energy band of the \rosat\ HRI to that of the
ensemble of bright point sources seen in the HRI image we find that 
they are similar.  The absorption-corrected $0.1-2.4$ keV flux of 
the hard X-ray component is $\sim7\times10^{-13}$ ergs cm$^{-2}$ 
s$^{-1}$ (Table 4).
The HRI flux from the three brightest point sources (having a 
combined count rate of 0.009 counts s$^{-1}$) is 
$\sim 1-1.3\times10^{-12}$ ergs cm$^{-2}$ s$^{-1}$ for 
$N_{\rm H} = 2.6-7.2\times10^{20}$ cm$^{-2}$ (Table 3). 
The hard component seen with $ASCA$ could thus account for 
most of the flux from the bright point sources
in the HRI image.

An identification of the hard component with the three
bright point-like sources seen in the HRI bandpass ($\sim$ 1 keV) then
argues strongly against the hardest ($E \geq 3$ kev)
X-rays being dominated by emission from a  
heavily absorbed AGN (both on morphological grounds, and because a heavily
absorbed source would produce too few counts in the HRI band). The point-like
nature
of the HRI sources also argues against inverse Compton scattering from a 
diffuse
population of radio-synchrotron-emitting electrons. Instead, emission from
a population of high-mass X-ray binaries is most likely (as ZGW conclude
on independent grounds).

\subsection{Physical Parameters of the X-Ray Gas}

The joint \rosat\ plus \asca\ X-ray spectrum of Arp 299 requires a
multi-component
fit, as outlined in section 3.2.  In addition, the different morphologies
of the hard and soft PSPC images suggest that these spectral components
are dominant over different spatial regions in Arp 299.  The soft, cooler
X-ray gas appears to be more extended than the hard, hotter X-ray gas (see
Fig. 2).  In order to derive the physical parameters of the X-ray gas and
compare them with those predicted from models of a superwind, 
we must estimate the appropriate volumes for the soft and hard
spectral components.  In the calculations that follow, we assume the hard,
0.77 keV thermal plasma occupies a region of $1.3'$ radius (our extraction
region for the core spectrum), corresponding to a volume of
$5.8\times10^{68}$ cm$^{3}$.  The soft, 0.20 keV plasma is assumed to
occupy a volume of $3.5\times10^{69}$ cm$^{3}$, equivalent to the volume
enclosed by the ``total" spectrum described in section 3.2.3.  We have
divided this volume into core and halo components, as in section 3.2,
in order to derive gas parameters for the soft thermal component. The emission
integral (the volume integral of the density squared) for the soft component
is $1.6\times10^{64}$ cm$^{-3}$,
60\% of 
which is contained within a radius of $1.3'$, the boundary of the
extracted core spectrum.  If the volume filling factor of the X-ray gas
is $f$, where $f\le1$, the physical parameters of the soft thermal component
can be estimated:
 
\noindent
1. The mean core gas density is $4\times10^{-3}f^{-1/2}$ cm$^{-3}$.
The mean halo gas density is $1.5\times10^{-3}f^{-1/2}$ cm$^{-3}$.
 
\noindent
2. For T$=2.3\times10^{6}$K, the corresponding presssures are
$2.5\times10^{-12}f^{-1/2}$ dyne cm$^{-2}$ for the core gas and
$1\times10^{-12}f^{-1/2}$ dyne cm$^{-2}$ for the halo gas.
 
\noindent
3. The mass of the gas at T$=2.3\times10^{6}$K in the core is
$1.9\times10^{9}f^{1/2}$ M$_{\odot}$.  The mass of gas at this temperature
in the halo is $3.5\times10^{9}f^{1/2}$ M$_{\odot}$.
 
\noindent
4. The thermal energy contents of the core
 and halo components are
$2.2\times10^{57}f^{1/2}$ erg and $4\times10^{57}f^{1/2}$ erg,
respectively.
 
\noindent
5. The radiative cooling times are $2\times10^{8}f^{1/2}$ yrs and
$6\times10^{8}f^{1/2}$ yrs, for the core and halo gas, respectively.
Here, we have used t$_{cool}=3kT(\Lambda n)^{-1}$, where 
$\Lambda=3.5\times10^{-
23}$ erg cm$^3$ s$^{-1}$
(Sutherland \& Dopita 1993) for gas at solar abundances at these temperatures.
 
We can similarly calculate the the physical parameters for the second
MEKAL thermal plasma identified in the spectrum of Arp 299
by our joint \rosat\ plus \asca\ spectral fits.  For a volume integral of the
density squared of this second, hard thermal plasma at kT=0.77 keV
of $3.7\times10^{63}$ cm$^{-3}$, the physical parameters of this
hot gas are:
 
\noindent
1. The mean gas density is $2.5\times10^{-3}f^{-1/2}$ cm$^{-3}$.
 
\noindent
2. The corresponding pressure is $6.3\times10^{-12}f^{-1/2}$ dyne cm$^{-2}$.
 
\noindent
3. The mass of gas at T$=9\times10^{6}$K is $1.2\times10^{9}f^{1/2}$ M$_{\o
dot}$.

\noindent
4. The thermal energy content of the gas is $5.5\times10^{57}f^{1/2}$ erg.
 
\noindent
5. The radiative cooling time of the gas is $1.3\times10^{9}f^{1/2}$ yrs
for solar metallicity (Sutherland \& Dopita 1993).

\section{Optical Spectra}

The long slit optical spectra can provide a wealth of details on the
physical conditions of the warm (T$\sim 10^4$ K) ionized gas in the
Arp 299 system.  The temperature, density, pressures, and excitation
state of this gas can in turn help constrain models of a starburst-driven
outflow.  The May87 and Jan88 slits are shown superimposed on the
H$\alpha$ image of Arp 299 in Fig. 10. The overall size of the
emission-line nebula is approximately 20 kpc.

As an
illustration of the diversity in the emission--line properties
seen across the Arp 299 nebula, we
have extracted seven, one-dimensional spectra covering the $6300\AA - 7000\AA$
wavelength range
from various locations 
in the nebula which
we feel are representative of the data set as a whole.  All seven 
spectra are from the medium resolution May87 data set, and are
pesented in Figs. 11a-g.  
The first two spectra
(Fig. 11a, b) are centered on components ``B1" and ``C", 
respectively, while the third spectrum (Fig. 11c) is from a 
region approximately $2''$ north of component ``A".  In Figs. 11a-c
the spectra 
have been generated by summing $4.5''$ along the appropriate slit.
The fourth spectrum (Fig. 11d) has
been generated by summing along a region $10''$ in length located $11''$ east
of component ``C" in the May87 longslit data.
Figures 11e, f, and g are spectra extracted from regions well
removed from the galaxy nuclei, and are chosen to show the strong [NII],
wide lines, and H$\alpha$ line splittings which characterize the faint gas
in the outer portions of the Arp 299 system (see the figure caption for a
description of the location of each of these spectra in the nebula).

In the following sections we describe the
physical state of the optical emission-line gas 
surrounding the two galaxy nuclei.

\subsection {Gas Densities and Pressures}

We have used the medium resolution May87 data to calculate the electron
density as a function of radius from components A, B, and C by measuring
the [SII]$\lambda\lambda$6717/6731 line flux ratio. For the Western half
of the emission-line nebula,
the geometric center of
sources B
and C is used to calculate the radial distances for the 
plot of density vs. radius shown in figure 12.
Although there are large errors associated with some of the
points beyond 1 kpc from the nuclei, it is clear from this plot that
the density drops with radius from a
peak of about 250 cm$^{-3}$ to the low density limit
($\approx50 $cm$^{-3}$) at radii of a few kpc.

Assuming that the temperature in SII zone is roughly $10^4$ K
(as appropriate for either photoionized or shock-heated gas - cf.
Dopita \& Sutherland 1996), the corresponding thermal pressures
are  P/k $\sim5\times10^6$ K cm$^{-3}$ in the inner region dropping
below P/k $\sim1\times10^6$ K cm$^{-3}$ beyond a radius of a few kpc.

\subsection {Extinction}

From the low resolution Jan88 data, we are able to measure the
Balmer decrement as a function of position along both slits.  These
measurements are converted to extinction measurements at H$\beta$ assuming
an intrinsic H$\alpha$-to-H$\beta$ line flux ratio of 2.86 and the
interstellar extinction curves of Mathis (1990).  Along the PA=90 slit,
the extinction peaks at nucleus A, yet is at a relative minimum to the
south of C (recall that this slit does not pass directly through component
C).  At the location of component A, H$\alpha$/H$\beta = 7.1$,
corresponding to 1.9 mag of extinction at H$\alpha$.  However, 2.3$''$
south of C the extinction at H$\alpha$ is only about 0.3 mag.  Between nuclei
A and B,
the extinction is approximately $1.2-1.4$ mag at
H$\alpha$.  Along the PA=120 slit, to the northwest of nucleus B, the
H$\alpha$-to-H$\beta$ line flux ratios imply extinctions
at H$\alpha$ of only $0.2-0.7$ mag. 
Dust is obscuring the gas
directly east of component A where the extinction at H$\alpha$ is greater
than 1.3 mag.  Although component A has the largest amount of foreground
extinction in the Arp 299 system, the regions between A and C (at the
interface of the two interacting galaxies) also appear to be heavily
reddened, with extinctions at H$\alpha$ of $0.7-1.3$ magnitudes, steadily
decreasing westward from component A.  The PA=120 Jan88 spectrum shows
that the outer regions of the system have relatively low extinction
compared to the nuclei, with less than 0.7 magnitude at H$\alpha$ at radii
12 to 20 arcsec northwest of component C.

Gehrz, Sramek and Weedman (1983) measure an H$\alpha$/H$\beta$ line flux
ratio of 5.5 at the position of component C through a 3.6$''$ aperture,
which is equivalent to an extinction of 1.4 mag at H$\alpha$.  Stanford \&
Wood (1989) measure HI column densities of $8.6\times10^{21}$ and
$3.1\times10^{21}$
cm$^{-2}$ respectively at the locations of components A and C, corresponding to
extinctions at H$\alpha$ of about 4.3 mag and 1.5 mag, respectively.
Although the lower values we see for extinction near component C as
compared to Gehrz et al.  and Stanford \& Wood can be explained if the
reddening is strongly peaked at C, the differences between our measured
value at A and the HI results cannot be explained this way since our slit
passed directly over component A.  The neutral hydrogen column densities
implied by the Balmer decrement are about half as large as
those measured by Stanford \& Wood. This disagreement is not surprising, since
both estimates depend upon uncertain assumptions:
the optically-derived columns assume a normal Galactic dust-to-gas
ratio and the radio-derived columns are directly proportional to the
assumed HI spin temperature.

We conclude that the extinction in Arp 299 is quite inhomogeneous. For
the purposes of the discussion to follow we will adopt a net extinction
at H$\alpha$ of 1 magnitude for the emission-line nebula. This will
doubtless underestimate the extinction deep inside the dusty starburst
regions (e.g. nucleus A), but we are primarily interested
in the energetic requirements of the large-scale nebula. The measured
H$\alpha$+[NII]$\lambda\lambda$6548,6584 flux of the nebula in Armus,
Heckman, \& Miley (1990)
then implies an extinction-corrected H$\alpha$+[NII]
luminosity of 5.3 $\times 10^{42}$ erg s$^{-1}$. Based on the typical
values measured for the ratio of [NII]$\lambda$6584/H$\alpha$ in the nebula
($\sim$ 0.4 to 0.5), the H$\alpha$
luminosity alone would be about 3.3 $\times 10^{42}$ erg s$^{-1}$.



\subsection {Emission Line Flux Ratios and Excitation}

Emission-line flux ratios, such as [OI]$\lambda$6300/H$\alpha$,
[NII]$\lambda$6584/H$\alpha$, [SII]$\lambda$6717+6731/H$\alpha$ and
[OIII]$\lambda$5007/H$\beta$, can provide reddening-insensitive
diagnostics of the of the ionizing sources in the Arp 299 system.  The
presence of the three strong ``nuclear" components (A,B, and C) along with
numerous other, fainter knots which are presumably HII regions powered by
young stars, are responsible for large variations of the emission-line
flux ratios with position along the spectral slits in the May87 and Jan88
data.  Among these data however, several trends are clear.  First, the
low-ionization
emission line flux ratios are generally at a local minimum at the
positions of the nuclei.  This is most evident in the [OI]/H$\alpha$
ratio, which originates in the partially ionized region where ionized and
neutral species co-exist in a bath of thermal electrons.  At the position
of component C, log [OI]/H$\alpha\sim-1.8$ and log
[NII]/H$\alpha\sim-0.7$, both of which are consistent with photoionization
by OB stars.  Second, off the nuclei, and away from isolated HII regions, the
line ratios can be much greater, reaching values of log
[OI]/H$\alpha>-1.0$ and log [NII]/H$\alpha\sim0$. The line ratios at various
locations are summarized in Table 5.

The trends noted above can be seen more clearly if one looks at a
plot of emission-line flux ratio vs. emission--line (in this case
H$\alpha$) surface brightness for the off--nuclear points, as is
illustrated in Fig. 13. Excluding the regions close to the nuclei,
the trend of increasing [NII]/H$\alpha$ and
[SII]/H$\alpha$ with decreasing nebular surface brightness is evident.
In contrast the [OIII]$\lambda$5007/H$\beta$ ratio is roughly constant at
values of $\sim$ 0.7 to 1.5 and shows no obvious trends with distance
from the nuclei or nebular surface brightness.

In addition to the correlation of the low--ionization emission--line flux ratios
with surface
brightness, there is also a correlation between the emission--line flux
ratios and the 
velocity dispersion of
the gas.  In Fig. 14 we plot
[NII]/H$\alpha$ and [SII]/H$\alpha$ flux ratios vs. H$\alpha$ linewidth in the
high resolution
May87 spectra.  As the line ratios increase, the velocity dispersion in
the gas also increases, suggesting that the physical and dynamical states
of the gas are coupled.
The
nuclear and near-nuclear points (those within $3''$ of the nuclei) do not
obey this correlation (see Fig. 14), presumably because the line-broadening
in these regions is largely gravitational in origin (e.g. we sample a 
significant piece of the rotation curve within the spectroscopic
sub-aperture).

In summary, the fainter, diffuse ionized gas in Arp 299 has both 
relatively stronger
low--ionization emission lines and is more kinematically disturbed than
the high-surface-brightness gas.

\subsection {Gas Kinematics}

With a number of slits covering the Arp 299 nebula, we can explore the
gas kinematics with position, and compare these with other measurements of
gas motions in the system, e.g. the HI measurements of Stanford \& Wood
(1989).  The ionized gas kinematics can be measured via the change in
emission-line centroids and widths as a function of position along the
slits.

The peak velocity of the H$\alpha$ line shows variations of $100-300$ km
s$^{-1}$ across the Arp 299 nebula.  In general, the velocities are blueshifted
to the south and redshifted to the north, with respect to the systemic velocity
of $\sim3150$ km s$^{-1}$, determined by Stanford \& Wood from the
centroid of the large-scale HI rotation curve.  Overall, the H$\alpha$
kinematics follow the HI rotation curve.  However, near the nuclei the velocity
of the H$\alpha$ line is at a relative minimum, suggesting the gas in the
immediate vicinity of the nuclei is blueshifted by $\sim50-100$ km
s$^{-1}$ with respect to the systemic velocity. This might arise as 
consequence of a dusty outflow of gas in which the backside flow is
partially obscured.

The H$\alpha$ linewidths range from less than 50 km s$^{-1}$ to nearly
600 km s$^{-1}$ across the Arp 299 system.
As with the line centroids, spatial variations in the
linewidths on the order of 200 km s$^{-1}$ are seen on scales as small as
1.5 kpc.  Near the galactic nuclei, the linewidths reach a local maximium,
typically about $200-300$ km s$^{-1}$.  However, if one excludes the
points near the nuclei (see above), there is a clear trend of increasing
linewidth
with decreasing emission-line surface brightness, as depicted in Fig. 15.
The broadest lines in the Arp 299 system, those with FWHM $>350-400$ km
s$^{-1}$, are found in the lowest surface brightness, diffuse ionized
gas in the outer regions of the nebula.

\section{Discussion}

\subsection{The Optical Emission-Line Nebula}

\subsubsection{The Ionization Source}

From the extinction-corrected H$\alpha$+[NII] luminosity we estimated
in \S 4.2  above, we deduce that the the total luminosity of the optical
emission--line nebula (summing over all emission lines) is $\sim10^{44}$
erg s$^{-1}$.  This represents about 4\% of the bolometric luminosity of
Arp 299 and several hundred times the X-ray luminosity of the nebula.

The luminosity of the nebula is so large that we can exclude the
possibility that it is heated {\it primarily} by mechanical energy
supplied either by the starburst or the galaxy collision itself.  The rate
at which a starburst produces mechanical energy (in the form of supernovae
and stellar winds) is only about 1\% of the starburst's bolometric
luminosity (Leitherer \& Heckman 1995).  Even if all this mechanical
energy is used to heat the emission-line nebula, the resulting luminosity
will only be 25\% of the observed value. Similarly, the kinetic energy of
10$^{10}$ M$_{\sun}$ of interstellar gas moving at 300 km s$^{-1}$ is
$\sim10^{58}$ ergs.  If this energy is entirely dissipated and radiated
away by the emission-line nebula during one dynamical crossing-time
($\sim$ 10$^8$ years), the resulting heating rate is only
3$\times$10$^{42}$ erg s$^{-1}$ or 3\% of the observed value.

We conclude that the {\it primary} source of energy input to the optical
emission-line nebula is most likely to be the ionizing radiation that
escapes from the starburst complexes and propogates into the nebula. The
Leitherer \& Heckman (1995) models predict a luminosity in ionizing
radiation that is roughly 10\% of the bolometric luminosity of the
starburst, so that only about 40\% of the available radiation is required
to power the nebula.

This picture provides a simple and natural explanation for the inverse
correlation between the relative strength of the low-ionization lines and
the gas surface brightness discussed above (Fig. 13).  For individual gas
clouds of a given density, the higher the value of the local intensity of
the ionizing radiation field, the higher the resulting value of the
ionization parameter $U$ (defined to be the ratio of the densities of
ionizing photons and electrons within a photoionized gas cloud). Simple
ionization equilibrium arguments show that $U$ determines the ionization
state of the gas, and the H$\alpha$ surface brightness will be
proportional, through recombination, to the intensity of the ionizing
radiation field. Thus, the proportionality between $U$ and the H$\alpha$
surface brightness will naturally produce enhanced relative intensities of
low-ionization lines in the faint gas (see Wang, Heckman, \& Lehnert
1998).

We can examine this more quantitatively.  The H$\alpha$ luminosity of the
nebula (3.3$\times$10$^{42}$ implies an ionization rate of $Q \sim
2.4\times10^{54} s^{-1}$. If we assume that this arises equally in nuclei
``A" and ``B1", then the radial electron density profile discussed above (Figure
12), leads to a predicted ionization parameter $U \sim 10^{-3}$ for the
gas located at radii of one to a few kpc from the nuclei.  J. Sokolowski
(private communication) has kindly run $CLOUDY$ models of gas that is
photoionized by a starburst population of hot stars using the evolutionary
synthesis code discussed by Leitherer \& Heckman (1995). We use the models
in which the cosmically-abundant, refractory elements (e.g. Fe and Si) are
locked-up in dust grains (as in the case of diffuse clouds in our own
Galaxy). This depletion of gas-phase coolants has the effect of increasing
the equilibrium temperature in the photoionized gas and thereby increasing
the emissivity in the collisonally-excited forbidden lines with respect to
the Balmer recombination lines (Shields \& Kennicutt 1995).  For the above
value of $U$, the predicted line-ratios are [SII]$\lambda\lambda$
6716,6730/H$\alpha$ = 0.4, [NII]$\lambda$6583/H$\alpha$ = 0.6,
[OI]$\lambda$6300/H$\alpha$ = 0.03, and [OIII]$\lambda$5007/H$\beta$ =
1.6, in reasonable agreement with the typical observed values in the 
{\it high}-surface-brightness part of the nebula (Figures 11, 13, and 14; Table
5).

While photoionization clearly dominates the total energy input into the
nebula and the excitation near the nuclei, mechanical heating also plays
an important role, especially in the diffuse low-surface-brightness 
gas. Firstly, recall that the
estimated mechanical luminosity of the starburst is 25\% of the nebula's
emission-line luminosity.
Secondly, the good correlation we find between the
widths of the emission-lines and the line-ratios (Figure 14) suggests that
there is a connection between the dynamics of the gas and its physical
state (as expected for shock heating). Finally, shock-heating may be
required to explain the extreme line-ratios present in the very faint
outer regions where [OI]/H$\alpha$ = 10 to 20\% and [OIII]/H$\beta \sim 1$
(see Table 5).  Most likely, both photoionization and shocks play a role,
with the former being more globally significant, and the latter being
important in the most kinematically-disturbed (faint, outer) regions.

\subsubsection{The Pressure Profile}

One property of the emission-line nebula that can be most clearly ascribed
to the mechanical energy supplied by the starburst is the gas
density/pressure profile (Figure 12).  In the starburst--driven wind
models of Chevalier \& Clegg (1985), there is a region of radius r$_{*}$
(the starburst) inside which mass and kinetic energy are injected at a
constant rate per unit volume, resulting in a region of hot gas which
slowly expands through a sonic radius (located approximately at the radius
of the starburst) and is then transformed into a supersonic outflow.

The gas pressure inside the starburst is just the static, thermal pressure
of the hot fluid that has been produced by thermalization (shock-heating)
in high-speed collisions between material ejected by supernovae and the
ambient interstellar medium. This pressure is essentially determined by
the size of the starburst and the rate at which mass and kinetic energy is
injected. Since the sound-crossing time inside the starburst is much
shorter than the outflow time (e.g. since the flow is subsonic inside the
starburst itself), the pressure is roughly constant throughout this
central region.

Using the starburst models of Leitherer \& Heckman (1995) to relate the
rate at which the starburst injects mass and thermal/mechanical energy to
its bolometric luminosity, we expect the central pressure to be given
approximately by:

\begin{equation}
P_0/k = 3 \times 10^6 L_{bol,11} [r_{*}/kpc]^{-2}\ K\,cm^{-3}
\end{equation}

At radii much larger than the starburst radius, the pressure of
the outflow is dominated by the wind's ram pressure. This will
essentially be set by the rate at which the starburst injects
mass and energy into the outflow, and will drop like the inverse square
of distance from the starburst:

\begin{equation}
P(r)/k = 3 \times 10^6 L_{bol,11} [r/kpc]^{-2}\ K\,cm^{-3}
\end{equation}

where $r$ is the distance from the starburst (assuming $r \gg r_{*}$).

These predictions agree reasonably well with the observations of
Arp 299. Taking L$_{bol}$\, =\,3\,$\times$\,10$^{11}$\,L$_{\sun}$
as a rough estimate of bolometric luminosity for each of the two starbursts
 and r$_{*}$ = 1.3 kpc as their estimated radii,
the predicted value P/k = 5 $\times$ 10$^6$ K cm$^{-3}$
is the same as that derived from the temperatures and densities given in 
\S 4.1. Similarly, the predicted
value for P/k at a radius of 3 kpc is $\sim10^6$ K
cm$^{-3}$ (n$_e$ = 50 cm$^{-3}$), consistent with Figure 12.

\subsection{The X-Ray Nebula}

In principle, the X-ray nebula could arise as a significant fraction of the
gas initially in the disks
of two spiral galaxies is heated collisionally during the
galaxies' on-going merger. As discussed in \S 5.1 above, the maximum rate of
collisional heating in a merger will be of-order 3$\times10^{42}$ erg s$^{-1}$
, which
is about an order-of-magnitude larger than the observed X-ray luminosity.
The mass
of hot gas (7$\times$10$^{9}$ M$_{\odot}$) is also consistent with this
picture.

There are
however at least two problems with this model. First, the observed gas
temperatures are much too high in the inner region. That is,
cloud collision speeds appropriate to a galactic merger (300
to 500 km s$^{-1}$) correspond to postshock temperatures of 1.3 to 3.5
million K, while the observed temperature is
9 million K in the inner region. Second,
if X-ray nebulae in colliding/merging galaxies are directly produced
by the galactic collision, then we would expect that bright X-ray nebulae
would be a common feature of mergers of disk galaxies
(regardless of whether these mergers are undergoing a starburst).
Wang et al (1997) and Read \& Ponman (1998) have shown that this is {\it not}
the case. In the Wang et al
sample of 12 merging/colliding galaxies (including Arp 299), the only
strong soft X-ray emitters are also strong H$\alpha$+[NII] and far-IR emitters.
Put another way, strong soft X-ray emission is not a ubiquitous feature
of galaxy mergers (Read \& Ponman 1998), but instead appears to share a
a common power source with the optical line and far-IR emission (a starburst).
We therefore regard the most likely origin of the X-ray nebula in
Arp 299 to be a
superwind.

Evidence for outflowing gas in Arp 299 comes from the
optical spectroscopic data on the kinematically-disturbed faint
emission-line filaments and loops (see \S 4.4).
The presence of diffuse X-ray-emitting gas extending well outside the optical
boundaries of the galaxy and oriented roughly perpendicular to the
system's HI major axis is also
highly suggestive
of an outflow.
In order to quantitatively assess whether the soft X-ray emission
can indeed be
produced by gas heated by supernovae and stellar winds in the central
star-bursting complexes, we
will adopt a simple model that assumes that
mechanical energy is being continuously injected inside the starburst and used
to heat the
surrounding gas. This gas then flows out of the starburst
and emits X-rays (cf. Wang 1995). In such a model we need to specify
the heating rate and the mass-injection rate into the outflow. 
 
As noted earlier, the IR luminosity of Arp 299, 
when used in conjunction with the starburst models of 
Leitherer \& Heckman (1995), 
corresponds to a total
mechanical heating rate of about $2 \times 10^{43}$ erg s$^{-1}$.
We will assume that the gas in the starburst
is initially heated to a temperature
of about $9 \times 10^6$ K (the temperature of the hottest diffuse gas that we
detect in our X-ray data). For the heating rate given above,
conservation of energy then implies that gas is heated at a rate of
$\sim100$ M$_{\odot}$ per year (Chevalier \& Clegg 1985).
This is about an order of magnitude greater
than the predicted rate at which supernovae and stellar winds return mass to
the interstellar medium in this system. This implies that the
outflow is strongly ``mass-loaded'': most of the outflowing material
is ambient interstellar gas in and around the starburst that has been heated by the
supernovae
and stellar winds (see Suchkov et al 1996). In effect, each
supernova heats up, on average, about 200 M$_{\odot}$ of interstellar gas.
In this case, we would expect the average metallicity of the X-ray emitting
gas to be similar to that in the interstellar medium of Arp 299.

Since the predicted mechanical heating rate is about $10^2$ times the total
luminosity of the two MEKAL thermal components, we will
ignore the dynamical effect of radiative cooling in our simple
model. This neglect is further justified since the estimated
radiative cooling time of the X-ray gas (approximately a Gyr - see
\S 3.3 above) is much longer than the adiabatic expansion timescale
($\sim few \times 10^{7}$ years).
 
Following Chevalier \& Clegg (1985) and Wang
(1995) we therefore consider a spherically-symmetric wind with a
mass outflow rate of 100 M$_{\odot}$ per year which is ``fed'' by hot gas
that is injected at a uniform temperature of $9 \times 10^6$ K.
We will take the overall size of the energy injection region
(encompassing both the Eastern and Western starbursts) to be similar to
that of the bright X-ray emission in the HRI data (radius $\sim$ 4 kpc).
This is obviously an oversimplication, but might roughly represent the
effect of the collision between the two outflows produced by the
Eastern and Western starbursts, and the subsequent shocks and rethermalization
of the gas on these size-scales. 
In the absence of gravitational decelleration, the wind would reach
a terminal velocity of v$_{wind}$ = 800 km s$^{-1}$ (Chevalier \& Clegg 1985).
We assume the wind cools through adiabatic expansion while
also emitting soft X-rays.
The emissivity of gas in the 0.1 to 2.4 keV energy band is constant
to within about a factor of $\sim3$ as a function of temperature over
the range between $4 \times 10^5$ K to $5 \times 10^7$ K
($\simeq 2 \times 10^{-23}$
erg cm$^{3}$ s$^{-1}$ for solar metallicity).
This then allows
us to predict an X-ray luminosity from the outflow of $\simeq
4\times10^{41}$ erg s$^{-1}$ from within
a region equal in size to the X-ray halo (radius of $\sim$ 20 kpc).
This rough estimate compares reasonably well with the measured luminosity of
the two MEKAL X-ray components of $2 \times 10^{41}$ erg s$^{-1}$.
The total predicted mass
of the hot gas will be about $3 \times 10^{9}$ M$_{\odot}$ and the predicted
thermal plus kinetic
energy in the outflow will be $2 \times 10^{58}$ ergs. Again, these
predicted values agree reasonably well with the derived values for the sum
of the hot (0.77 keV) and warm (0.2 keV) components:
M $\simeq 7 \times 10^9 f^{1/2}$ M$_{\odot}$
and E$_{therm} = \simeq 10^{58} f^{1/2}$ ergs.

Note that a radius of 20 kpc
and a wind terminal velocity of 800 km s$^{-1}$ corresponds to a rough
dynamical age of 25 Myr (reasonable for a starburst). If the outflow
has lasted much longer than this, the total mass and energy in the flow
will rise accordingly, but this will comprise
material at larger radii ($>$ 20 kpc). Such material will have a low
density and (due to adiabatic cooling) a low temperature, and would
therefore not produce significant X-ray emission (cf. Wang 1995).
In fact, our simple model predicts a temperature range in the hot gas
ranging from $9 \times 10^6$ K in the energy injection region to 
$6 \times 10^6$ K at a radius of 4 kpc to $5\times 10^5$ K at a radius
of 20 kpc. These temperatures then span the observed/inferred range,
and the temperatures reach the lower limit where the ROSAT PSPC sensitivity 
plummets ($T \leq 4\times10^5$ K) at roughly the same radius as where
the observed emission ends. 
 
Having demonstrated that this simple model can very roughly reproduce the gross
properties of the X-ray emission, we next consider the fate
of this outflowing gas. Following Wang (1995), the ``escape
temperature'' for hot gas in a galaxy potential with an escape
velocity v$_{es}$ is given by:
 
T$_{es}$ $\simeq$ $9.9\times 10^{5}$ (v$_{es}$/300 km s$^{-1}$)$^{2}$ $ ^o$K
 
The HI rotation curve of Arp 299
has an amplitude of 240 km s$^{-1}$ out to the edge
of the HI distribution at a radius of 22 kpc (Stanford \& Wood 1989).
Assuming that Arp 299 has an
isothermal dark matter halo extending to a radius of 300 kpc, the
escape velocity (see equation 13 in Heckman et al. (1995))
at r = 4 (22) kpc is about 780 (650) km s$^{-1}$ and
T$_{es}$ = 6.7 (4.6) $\times 10^6$ K.
These temperatures
are similar to the range of temperatures we measure for the
hot gas ($\simeq 2 \times 10^6$ K for the cooler component and
$\simeq 9 \times 10^6$ K for
the hotter component). The escape velocity is also similar to the estimated
terminal velocity for the wind (see above). These rough arguments
suggest that the outflowing gas {\it may} be
able to escape the gravitational potential of Arp 299.

\section{Conclusions}

We have discussed optical and X-ray data on the spectacular gaseous
nebula surrounding the infrared-luminous merger/starburst system
Arp 299 (= NGC 3690 or Mrk 171).

We suggest that the on--going galaxy collision has tidally redistributed
the cold interstellar media of the merging galaxies. 
The total luminosity of the optical emission-line nebula
(10$^{44}$ erg s$^{-1}$ $\sim$ 4\% L$_{bol}$) is too high 
to be powered primarily by either the collision of two galaxies directly, 
or by a starburst-driven superwind. Instead, the gas is mostly photoionized
by radiation from the ongoing starburst.

We find strong trends in which the relative strengths of low-ionization
lines ([SII], [NII], and [OI]) increase systematically with both decreasing
nebular surface brightness and with increasing velocity dispersion. 
These trends can be understood if both hot stars and large--scale shocks
play a role in energizing the gas.  The shocks become increasingly
important in the faint, outer regions of the nebula, and they provide
for the coupling between the dynamical and physical state of the gas.

We also find that the gas pressures are high in the inner optical emission-line
nebula (P/k $\sim$ 5$\times$10$^6$ K cm$^{-3}$ within a radius of a kpc,
and fall systematically with radius (dropping below P/k $\sim$ 10$^6$ K
cm$^{-3}$ at a radius of $\sim$ 3 kpc). The shape and amplitude of the
radial pressure profile agrees with the pressures predicted for
starburst-driven outflows.

While the luminosity ($\sim 2\times10^{41}$ erg s$^{-1}$) and mass ($\sim
7\times10^9$ M$_{\odot}$) of the X-ray nebula can be explained by the
collision of the interstellar media of two galaxies, we believe a more
plausible heating source is the superwind.  The thermal energy content
($\sim10^{58}$ ergs) and dynamical age ($\sim few\times10^7$ years) of the
X-ray nebula agree with a superwind model, but its high gas mass and X-ray
luminosity, and it relatively low temperature (9 million K in the inner
region and 2.3 million K in the outer) imply that the outflow must be
strongly `mass-loaded' (most of the X-ray emitting gas is ambient
interstellar material that has been mixed into the outflow).  The implied
outflow rate of $\sim 100 M_{\odot}$ per year can be compared to the estimated
star-formation rate in the starburst of 40 (100) $M_{\odot}$ per year for
a Salpeter IMF extending down to 1 (0.1) $M_{\odot}$.  The inner gas
temperature of 9 million K implies a terminal velocity for such a flow of
about 800 km s$^{-1}$. These respective values are similar to the
estimated escape temperature and escape velocity from the system.

Thus, perhaps the most striking result of this investigation is that in
Arp 299 a powerful starburst has apparently been able to heat up a
significant fraction of the interstellar medium of the merger product
(nearly 10$^{10}$ M$_{\sun}$) to a temperature and outflow speed that
may allow this gas ultimately to escape the system. This has implications
for understanding how disk-disk galaxy mergers may produce (at least some)
elliptical galaxies (Mihos \& Hernquist 1994; Schweizer 1992) and how the
intergalactic medium could have been polluted by metal-enriched material
(see Gibson, Loewenstein, \& Mushotzky 1997).

{\bf Acknowledgments}

This research was primarily supported by NASA LTSA grants NAGW-3138
to TH and NAGW-4025 to KW.
The research by LA described in this paper was carried out by the California
Institute of Technology, under a contract with NASA.
This
research has made use of the NASA/IPAC extragalactic
database (NED), which is operated by the Jet Propulsion Laboratory, Caltech,
under contract with the National Aeronautics and Space Administration.

\clearpage
\onecolumn

\begin{table}
\caption [ ] {Spatially-Resolved PSPC Spectra}
\begin{tabular}{lccccccc}
\hline
Region & Geom. & \multispan{2}{}{~~Extent~~} & Orientation & Area & Rate & Counts$^{b}$ \\
       & Shape & Maj. ax. $a [~']$ & Min. ax. $b [~']$ & PA $[~^{\circ}]$ & [arcmin$^2$] & [c/s] &  \\
\hline
\hline
{\bf Total}   & Ellipse            & 2.78 & 2.22 & 193 & 19.38 & 0.099 & 983 \\
{\bf Core} & Circle             & 1.3  & 1.3  &     &  5.31 & 0.087 & 863 \\
{\bf Halo}    & Ellipse $-$ Circle &      &      &     & 14.07 & 0.012 & 119 \\
\hline
\end{tabular}
\tablenotetext{}{{\bf Notes to Table 1}:}
\tablenotetext{a}{Regions used to extract the PSPC spectra 
discussed in $\S3.2$.  The halo region is bounded by the 
elliptical region used to extract the total spectrum and the 
circular region used to extract the core spectrum.} 
\tablenotetext{b}{Total number of source counts.}
\end{table}
\clearpage
\onecolumn

\begin{table}
\caption [ ] {Results of PSPC Spectral Modelling}
\begin{tabular}{lcrrccr}
\hline
Region$^{a}$ & Model$^{b}$ & N$_{\rm H}^{c}$ & kT$_{\rm M}^{d}$ & Z$^{e}$/Z$_{\odot}$ 
& $\Gamma$ or (kT$_{\rm B}^{f}$)& $({\chi^2}/{\nu})^{g}$\\ 
\hline
\hline
H & B & 0.9(f) & 0.20$^{+0.12}_{-0.07}$ & 1.0 &\nix & \nix \\
C & A & 7.4$^{+2.6}_{-1.6}$ & \nix & \nix & 2.4$\pm$0.3 & 83/33 \\
C & B1 & 0.9(f) & 0.83$^{+0.10}_{-0.04}$ & 1.0 &\nix & 66/34 \\
C & B2 & 1.2$^{+0.6}_{-0.4}$ & 0.83$^{+0.16}_{-0.09}$ & 1.0 
   & \nix& 63/33 \\
C & B3 & 4.0$^{+1.6}_{-1.4}$ & 0.85$^{+0.15}_{-0.13}$ & 0.12$^{+0.28}_{-0.07}$
   & \nix & 41/32 \\ 
C & C & 2.2$^{+0.9}_{-0.6}$ & 0.70$^{+0.14}_{-0.18}$ & 1.0 &  
   7.0(f)& 29/32 \\ 
\hline
\end{tabular}
\tablenotetext{}{{\bf Notes to Table 2}:}
\tablenotetext{a}{The spectral extraction region; H=halo and C=core; f=fixed
parameter.}
\tablenotetext{b}{Models are A: absorbed power-law,
B: MEKAL plasma with absorption ($N_{\rm H}$(Gal) = $0.9 \times 
10^{20}$ cm$^{-2}$), C:
MEKAL plasma and thermal Bremsstrahlung with absorption.}
\tablenotetext{c}{The absorbing column density in units of 10$^{20}$ cm$^{-2}$.} 
\tablenotetext{d}{The MEKAL temperature in units of keV.}
\tablenotetext{e}{MEKAL plasma abundance in solar units.} 
\tablenotetext{f}{The photon index (or temperature) of the 
hard X-ray power law (or Bremsstrahlung component in units of keV.)} 
\tablenotetext{g}{Value of $\chi^2$ divided by the number of degrees of freedom.}
\tablenotetext{}{Errors are $\chi^2_{\rm min}$ + 4.61 or 6.25,
which are 90\% confidence for 2 or 3 free parameters.}
\end{table}

\begin{table}
\caption [ ] {Results of joint PSPC and $ASCA$ fits}
\small
\begin{tabular}{lccccccc}
\hline
Model$^a$&kT$_{\rm M2}^b$& ${Z/Z_{\odot}}^c$ & $A_{\rm M2}^d$ & N$_{\rm H}^e$
    &$\Gamma^f$ or (kT$_{\rm B}^f$) & $A_{\rm H}^g$ & $\chi^2/\nu^{h}$ ($\chi_{\
nu}^2$)\\
    & [keV] &  & [$10^{-3}$] & [$10^{20}$] &  & [$10^{-4}$] &  \\
\hline
\hline
A & \nix & \nix & \nix & 5.6$\pm$1.0 & 2.27$\pm$0.11
    & 4.3$\pm$0.2 & 439/279 (1.57) \\
B & 2.7$^{+0.4}_{-0.3}$ & 0.07$^{+0.13}_{-0.07}$ & 1.4$\pm$0.2
    & 2.8$\pm$0.5 & \nix & \nix & 537/278 (1.93) \\
C & 0.70$^{+0.12}_{-0.07}$ & 0.27$^{+0.38}_{-0.21}$
    & 0.68$^{+0.74}_{-0.58}$ & 2.6$\pm$0.7 & 1.51$^{+0.22}_{-0.27}$
    & 1.6$^{+0.4}_{-0.6}$ & 285/276 (1.03) \\
D1$^i$& 0.77$^{+0.08}_{-0.17}$&1.0(f)&0.18$^{+0.02}_{-0.04}$
    & 7.2$^{+6.7}_{-4.0}$ & 1.6$^{+0.15}_{-0.20}$/(11$^{+10}_{-4}$)
    & 2.1$^{+0.2}_{-0.3}$/(2.4) & 279/276 (1.01) \\
D2$^i$ & 0.77$^{+0.08}_{-0.17}$&1.0(f)&0.14$^{+0.02}_{-0.03}$
   & 6.2$^{+6.0}_{-3.7}$ & 1.6$^{+0.15}_{-0.20}$/(11$^{+10}_{-4}$)
   & 2.2$^{+0.2}_{-0.3}$/(2.5) & 281/276 (1.02) \\
\hline
\end{tabular}
\tablenotetext{}{\bf Notes to Table 3:}
\tablenotetext{a}{Models are A: absorbed power law, B: MEKAL with
absorption; C:MEKAL plus power law with absorption, and D:
2-temperature MEKAL plus hard component.}
\tablenotetext{b}{MEKAL temperature.}
\tablenotetext{c}{Plasma abundance.}
\tablenotetext{d}{MEKAL normalization in units of
$10^{-14}$/($4\pi D^2$) $\int n_e n_H dV$, where $D$ is the
distance to the source (cm), and $n_e$ and $n_H$ are the electron 
and hydrogen densities (cm$^{-3}$)}
\tablenotetext{e}{Absorbing column density in units of cm$^{-2}$.}
\tablenotetext{f}{The photon index (or Bremsstrahlung
temperature in units of keV).}
\tablenotetext{g}{Normalization of the hard component.  The power-law
normalization is in units of photons keV$^{-1}$ cm$^{-2}$ s$^{-1}$
at 1 keV; the bremsstrahlung normalization is in units of
$3.02\times10^{-15}$/($4\pi D^2$) $\int n_e n_i dV$,
where $n_e$ is the electron density (cm$^{-3}$), $n_i$ is the ion
density (cm$^{-3}$), and $D$ is the distance to the source (cm).}
\tablenotetext{h}{Value for $\chi^2$ divided by the
number of degrees of freedom (reduced $\chi^2$).}
\tablenotetext{i}{The soft component (not listed) has kT = 0.2 keV
 and \NH\ = $0.9\times10^{20}$ cm$^{-2}$.  For model D1, the intrinsic
 absorption (column 5) is applied to the medium and hard
 components.  For model D2, the intrinsic absorption is only
 applied to the hard component.  The MEKAL normalization of the soft
 component is $0.84^{+0.32}_{-0.31}$ (D1) and
 $0.51^{+0.21}_{-0.20}$ (D2); same units as column 4.}
\end{table}

\begin{table}
\caption [ ] {Absorption-corrected Fluxes from the \rosat\ PSPC and \asca\ fits}
\begin{tabular}{lccccc}
\hline
Model&F$_{(0.1-2.4){\rm M1}}^{a}$&F$_{(0.1-2.4){\rm M2}}^{b}$
&F$_{(2-10){\rm M2}}^{c}$&F$_{(2-10){\rm Hard}}^{d}$
& F$_{(0.1-2.4){\rm Hard}}^{e}$ \\
\hline
\hline
C & \nix & 7.0$\pm$0.4 & 0.18$\pm$0.03 & 11.0$\pm$0.2 & 9.0$\pm$0.4 \\
D1 & 2.3$\pm$0.6 & 5.6$\pm$0.8 & 0.23$\pm$0.09 & 10.0$\pm$0.2 & 6.6$\pm$0.3 \\
D2 & 1.4$\pm$0.5 & 4.5$\pm$0.6 & 0.18$\pm$0.08 & 10.7$\pm$0.3 & 6.6$\pm$0.3 \\ 
\hline
\end{tabular}
\tablenotetext{}{{\bf Notes to Table 4}:}
\tablenotetext{}{Fluxes are given in units of $10^{-13}$ ergs 
cm$^{-2}$ s$^{-1}$. The models in column 1 correspond to 
those described in Table 3.}
\tablenotetext{a}{The $0.1-2.4$ keV flux for the 
0.2 keV MEKAL component, derived from the PSPC.}
\tablenotetext{b}{The $0.1-2.4$ keV flux 
for the $\sim$0.75 keV MEKAL component, derived from the PSPC.}
\tablenotetext{c}{The $2-10$ keV flux for 
the $\sim$0.75 keV MEKAL component, derived from \asca.
\asca\ fluxes are instrument-averaged fluxes.}
\tablenotetext{d}{The $2-10$ keV flux 
for the hard X-ray component, derived from \asca.}
\tablenotetext{e}{The $0.1-2.4$ keV flux
for the hard X-ray component, derived from \asca.}
\end{table}

\begin{table}
\caption [ ] {Optical Emission-Line Ratios in Arp 299}
\begin{tabular}{lccccc}
\hline
Location & \multicolumn{5}{c}{Log of Line Ratio} \\
&[OIII]$\lambda5007$/H$\beta$ & H$\alpha$/H$\beta$ & [OI]$\lambda6300$/H$\alpha$
& [NII]$\lambda6584$/H$\alpha$ & [SII]$\lambda6724$/H$\alpha$ \\
\hline
\hline
A & $-0.05$ & 0.80 & $-1.08$ & $-0.34$ & $-0.37$ \\
B & \nix & \nix & $-1.24$ & $-0.34$ & $-0.50$ \\
C & 0.00 & 0.53 & $-1.75$ & $-0.47$ & $-0.72$ \\
HSB NEB & $-0.1$ & 0.57 & $-1.6$ & $-0.6$ & $-0.6$ \\
	& 0.0 & 0.80 & $-1.0$ & $-0.3$ & $-0.3$ \\
LSB NEB & $-0.1$ & 0.46 & $-1.3$ & $-0.6$ & $-0.5$ \\
	& 0.1 & 0.60 & $-0.6$ & $-0.2$ & $-0.1$ \\
\hline
\end{tabular}
\tablenotetext{}{{\bf Notes to Table 5}:}
\tablenotetext{}{
We adopt the nomenclature of Gehrz, Sramek, \& Weedman (1983)
wherein source A corresponds to the nucleus of the Eastern galaxy, source B
corresponds
to the nucleus of Western galaxy, and source C is another site of intense
activity in this galaxy. These values are taken from Armus, Heckman, \& Miley
(1989).}
\tablenotetext{}{We divide the off-nuclear regions of the Arp 299
into regions of high- and low-H$\alpha$ surface-brightness, since the
properties of the emission-line gas correlate well with this parameter
(see Figures 13 and 15 and text for details). For both regions we list
the range of typical values for these line ratios, from minimum (top line)
to maximum (bottom line). See Figure 13 and the text.
}
\end{table}

\onecolumn

\clearpage
\vfill\eject
  
\clearpage

\begin{figure}
\figurenum{1}
\caption{\rosat\ PSPC $0.1-2.4$ keV image of Arp 299 
superimposed on the digital sky-survey
optical image.  The X-ray data have been blocked by a factor of three, 
and then smoothed with a Gaussian kernel having a $\sigma=5$ pixels 
($7.5''$).}
\end{figure}
\begin{figure}
\figurenum{2}
\caption{\rosat\ PSPC (a) hard-band and (b) soft-band contours
for Arp 299.  The X-ray images cover the energy bands of
$0.1-0.5$ and $0.5-2.4$ keV, respectively.  The data have been blocked
by a factor of three (as in Fig. 1)
and then smoothed with a Gaussian kernel having a $\sigma=10$ pixels
(15 arcseconds).
The contours in Fig. 2a are drawn at 90, 75, 50, 25, 10, 5, and 3\% of the 
peak value.  The contours in Fig. 2b are drawn at 90, 70, 60, 50, and 30\% 
of the peak value.  The lowest contour is is at the $\sim5\sigma$ level.
}
\end{figure}

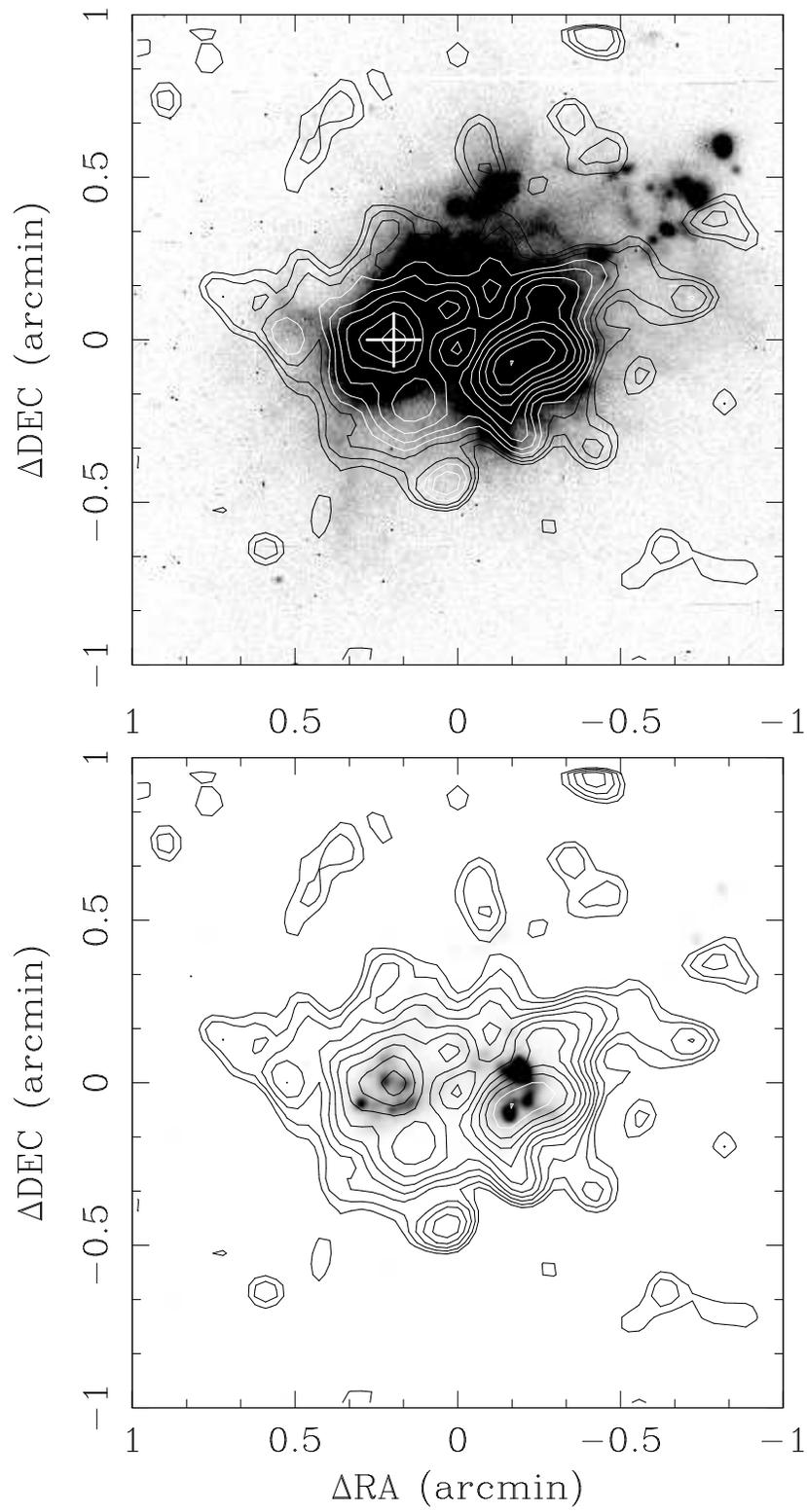
\begin{figure}
\figurenum{3} 
\caption{Two views of the \rosat\ HRI image overlayed on the continuum 
subtracted H$\alpha$+[NII] image from Armus, Heckman \& Miley (1990).  
In both frames the HRI data, shown as contours, have been blocked 
into 2$''$ pixels and
then smoothed with a Gaussian having a $\sigma=$1.5 pixels ($\sim$ 7$''$
FWHM). 
North is up and east is to the left in both frames.  The position of 
nucleus ``A" is indicated by a cross in the upper frame.
} 
\end{figure}


%
\begin{figure}
\figurenum{4}
\caption{\rosat\ PSPC spatially-resolved spectra of the core
and halo regions shown    
together for comparison. The sizes of the extraction regions  
are listed in Table 1.  Triangles indicate the 
halo spectrum; crosses indicate the core spectrum.
}
\end{figure}
%


%
\begin{figure}
\figurenum{5}
\caption{Ratio of the \rosat\ PSPC spectrum of the core region 
in Arp 299 to models consisting of 
(a) a power law (chosen to illustrate the
line emission near 0.9 keV), (b) a MEKAL plasma 
with free $N_{\rm H}$ and solar abundances (Model B2
in Table 2), (c) a MEKAL plasma plus thermal 
bremsstrahlung with free $N_{\rm H}$ and solar abundances
(Model D in Table 2). 
} 
\end{figure}
%


%
\begin{figure}
\figurenum{6}
\caption{Ratio of the \rosat\ PSPC and \asca\ data to a MEKAL plasma model  
with free abundances (model B in Table 3).
}
\end{figure}
%


%
\begin{figure}
\figurenum{7}
\caption{\rosat\ PSPC and \asca\ data and the best-fitting 
three-component model folded through the instrumental response  
(model D2 in Table 3). Data from all detectors are shown.
}
\end{figure}
%


%
\begin{figure}
\figurenum{8}
\caption{The three-component model that best describes
the $\sim0.1-10$ keV X-ray spectrum of
Arp 299.  Here, extra $N_H$ is applied to both the medium and
hard components (Model D2).
}
\end{figure}
%


%
\begin{figure}
\figurenum{9}
\caption{The ratio of the \rosat\ PSPC and \asca\ data to the best-fitting
three-component model (model D2 in Table 3).
}
\end{figure}
%


%
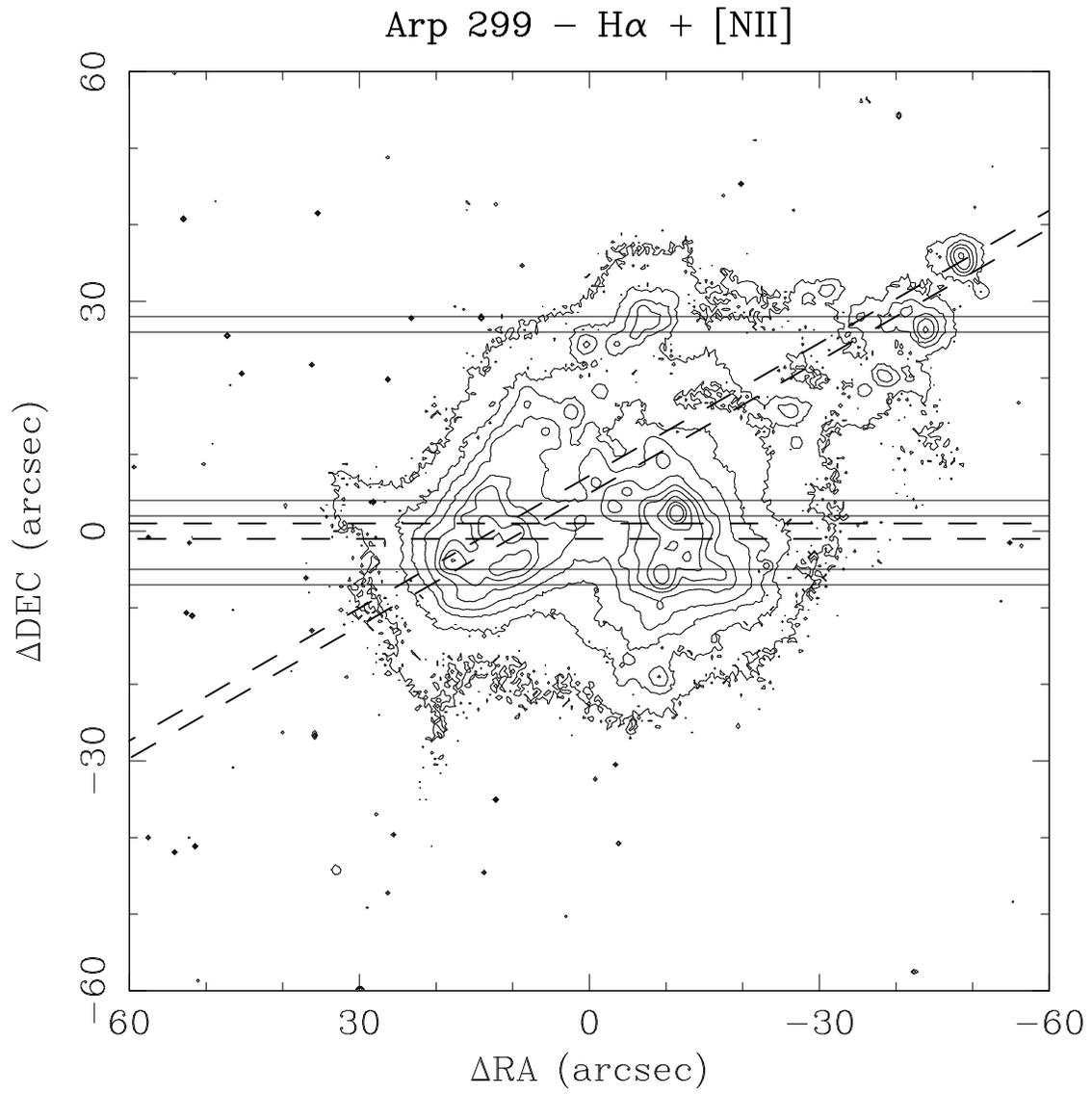
\begin{figure}
\figurenum{10}
\caption{Continuum subtracted H$\alpha+$[NII] image of Arp 299 showing 
the positions of the spectroscopic slits discussed in the text.  The slits
used to obtain the medium resolution May87 data are indicated by solid lines,
while the slits used to obtain the low resolution Jan88 data are indicated
by dashed lines.
}
\end{figure}
\begin{figure}
\figurenum{11}
\caption{Medium resolution spectra extracted from the May87 longslit data
at various positions throughout the Arp 299 nebula.  Spectra of components 
``B1" and ``C", are shown in Figs. 11a \& 11b, respectively.  Fig. 11c is from a
position, $2''$ north of component ``A".  Figs. 11a-c have been created 
by summing $4.5''$ along the slit.  The spectrum of the gas directly between
components ``A" and ``C" is shown in Fig. 11d, created by summing $10''$ 
along the slit, centered $11''$ east of component ``C".  Figs. 11e-g are 
spectra extracted from the faint, nebular gas as follows: (11e) $24''$ west
of ``C", summed over $32''$; (11f) $23''$ west of ``B1", summed over $13''$; 
(11g) $24''$ north and $6.5''$ east of ``A", summed over $23''$.  In all cases,
a relative flux density scale is plotted on the y-axis and wavelength, in $\AA$
is plotted along the x-axis.
See Fig. 10
for a visual representation of the slits on the H$\alpha+$[NII] image of Arp 299.}
\end{figure}

\begin{figure}
\figurenum{12}
\caption{Plot of the log of electron density (cm$^{-3}$),
as measured from the ratio of the 
[SII] lines in the May87 data, as a function of log of the distance 
(kpc) from the Arp 299 nuclei. For the western part of the nebula the
distance was measured with respect to the centroid of sources B and C,
while in the eastern part the distances are measured with respect to
source A. See text for details.
The shape and normalization of the radial density profile
(and hence the pressure profile) is 
as expected in the superwind model.
}
\end{figure}
\begin{figure}
\figurenum{13}
\caption{Plot of emission-line flux ratio vs. H$\alpha$ surface brightness 
(in arbitrary units) of
the emission-line gas in the Arp 299 nebula. Squares refer to the ratio of
[NII]$\lambda$6584/H$\alpha$ and triangles refer to [SII]$\lambda$$\lambda$
6717,6731/H$\alpha$. The open (solid) symbols refer to gas located at
radii $\leq$ ($>$) 3 arcsec from the galaxy nuclei.
Notice the trend of increasing 
[NII]/H$\alpha$ and [SII]/H$\alpha$ with decreasing surface brightness
for the off-nuclear regions (solid symbols).
}
\end{figure}
\begin{figure}
\figurenum{14}
\caption{The [NII]/H$\alpha$ (squares)
and [SII]/H$\alpha$ (triangles) flux ratios vs. the FWHM of the
the H$\alpha$ line (km s$^{-1}$). The open (solid) symbols refer to gas
located at
radii $\leq$ ($>$) 3 arcsec from the galaxy nuclei.
Notice the trend of increasing line-flux ratio with increasing H$\alpha$ 
line width in the off-nuclear gas
(solid symbols), consistent with the idea of shocks exciting
the faintest extra-nuclear 
gas and disturbing it kinematically.
}
\end{figure}
\begin{figure}
\figurenum{15}
\caption{Plot of the H$\alpha$ surface brightness (arbitrary units)
versus the the H$\alpha$ linewidth (FWHM in km s$^{-1}$).
The open (solid) symbols refer to gas
located at
radii $\leq$ ($>$) 3 arcsec from the galaxy nuclei.
The correlation of increasing 
linewidth and decreasing nebular surface brightness is clear
in the off-nuclear region (solid symbols).
}
\end{figure}

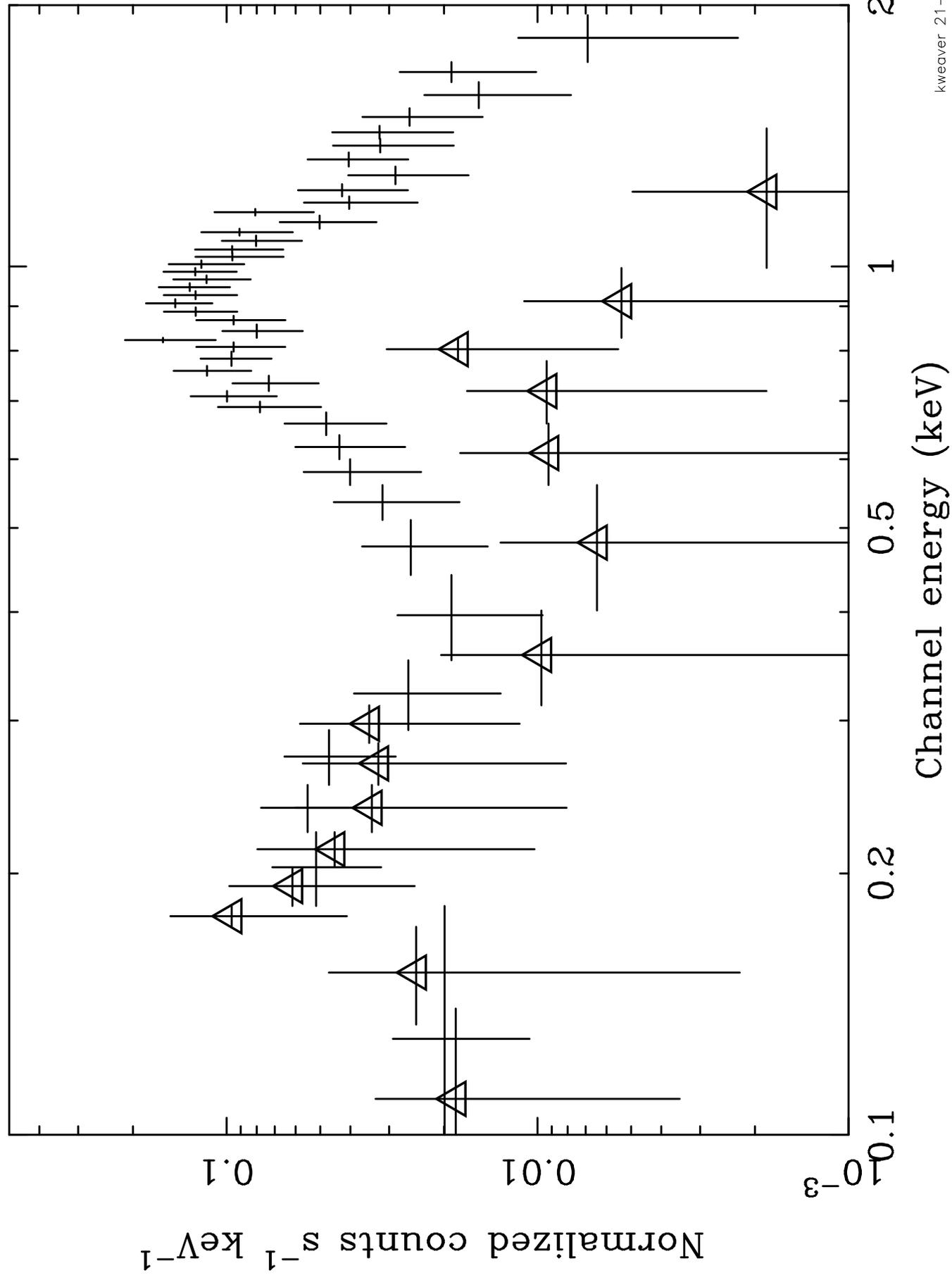
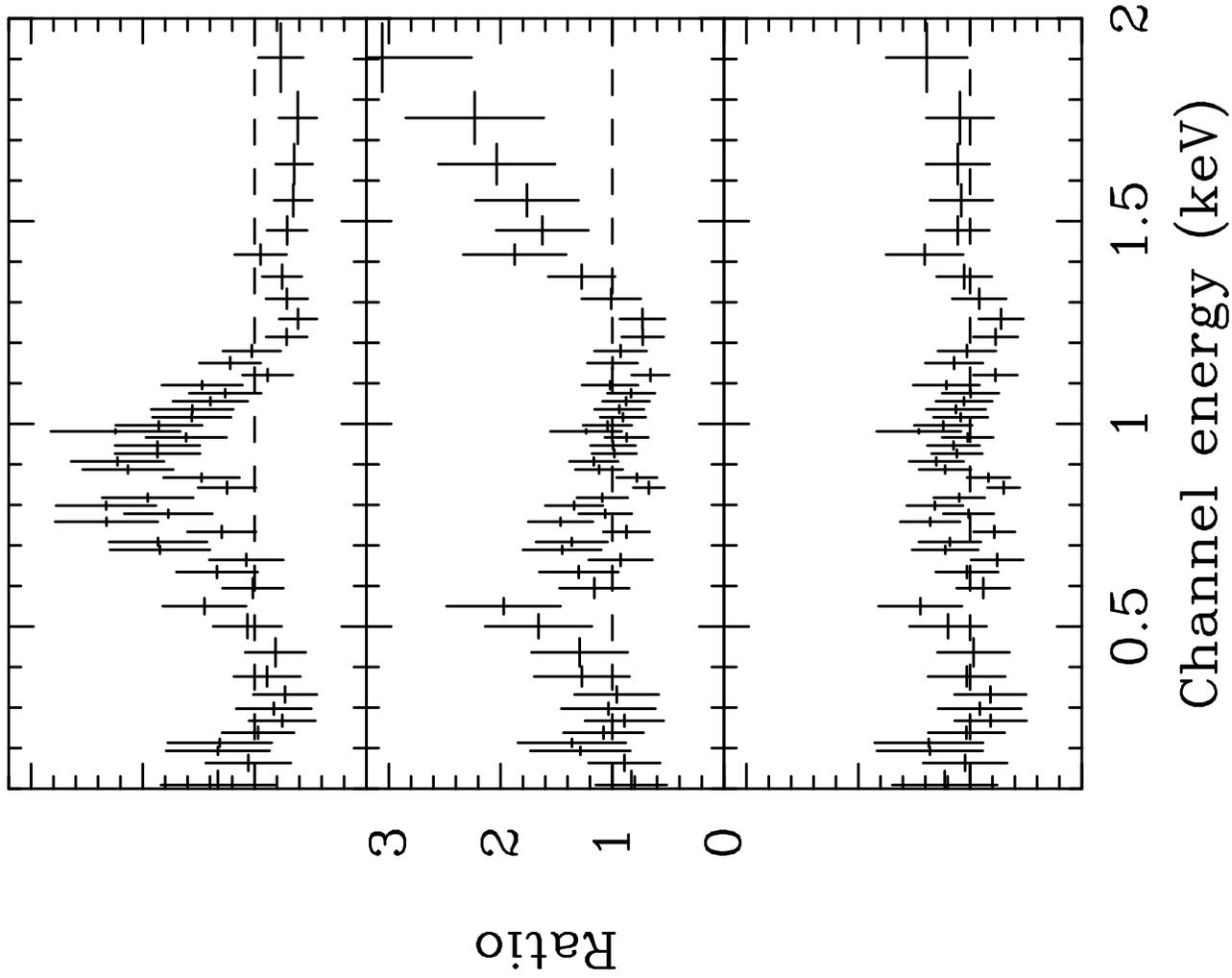
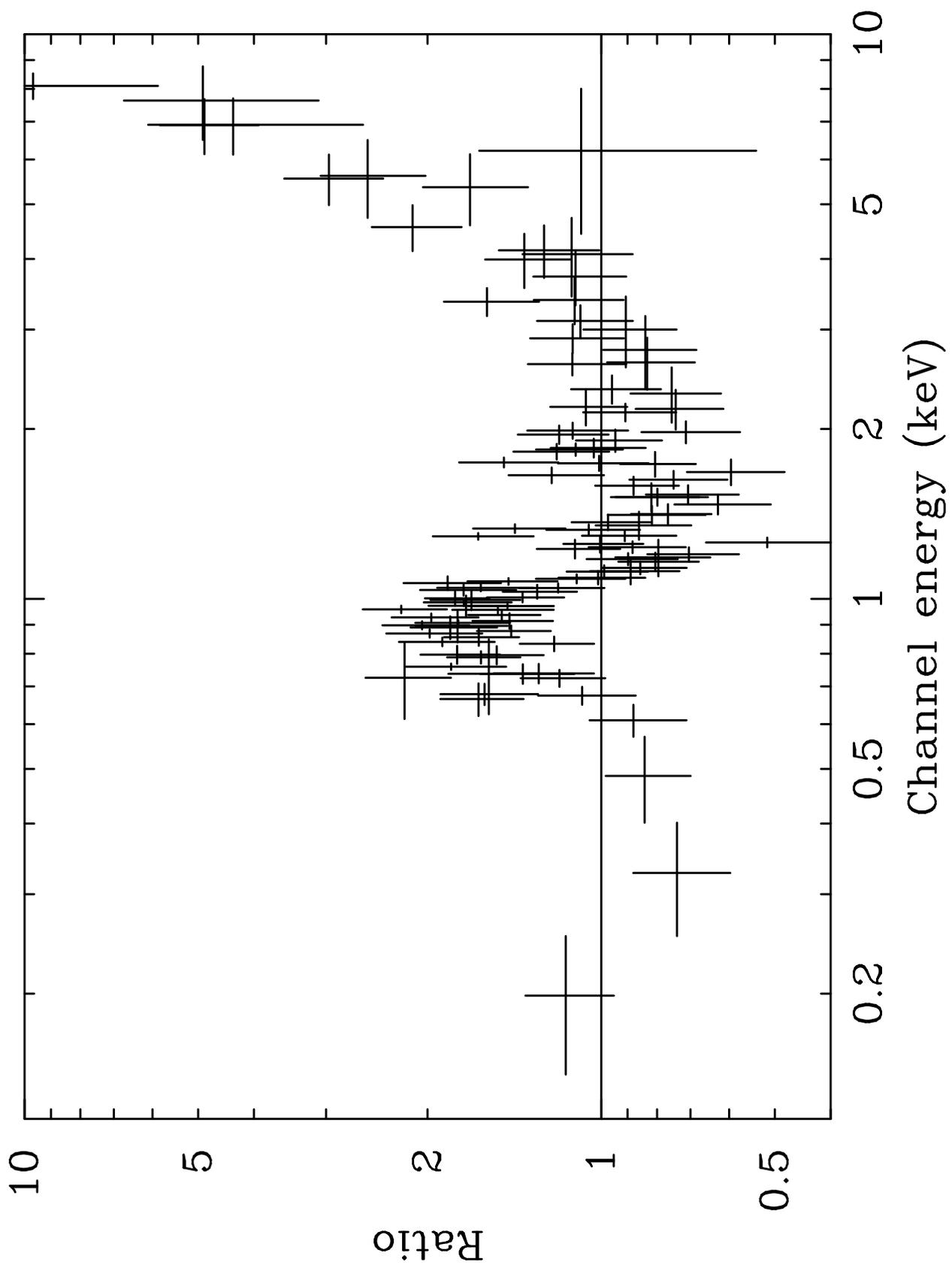
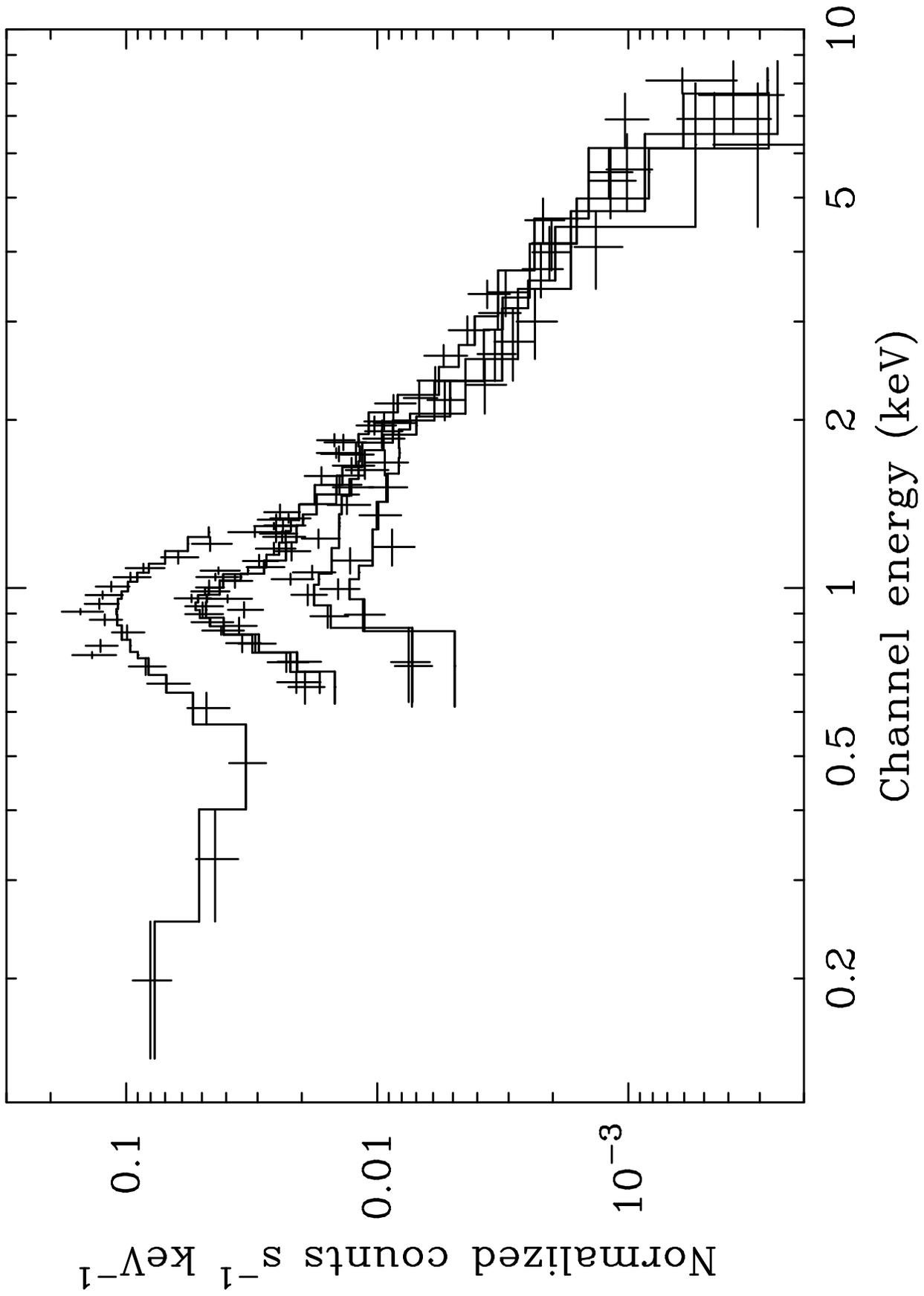
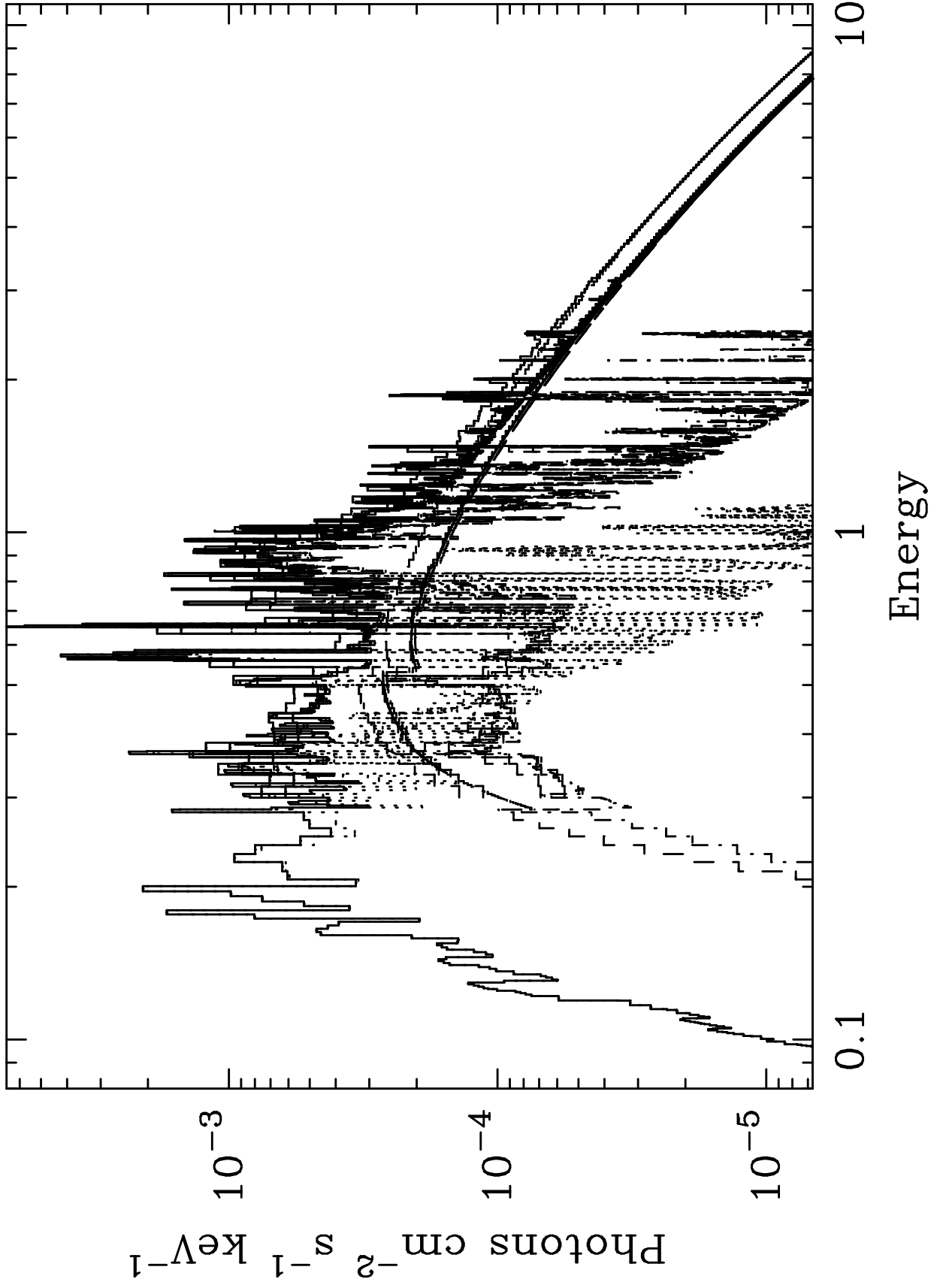
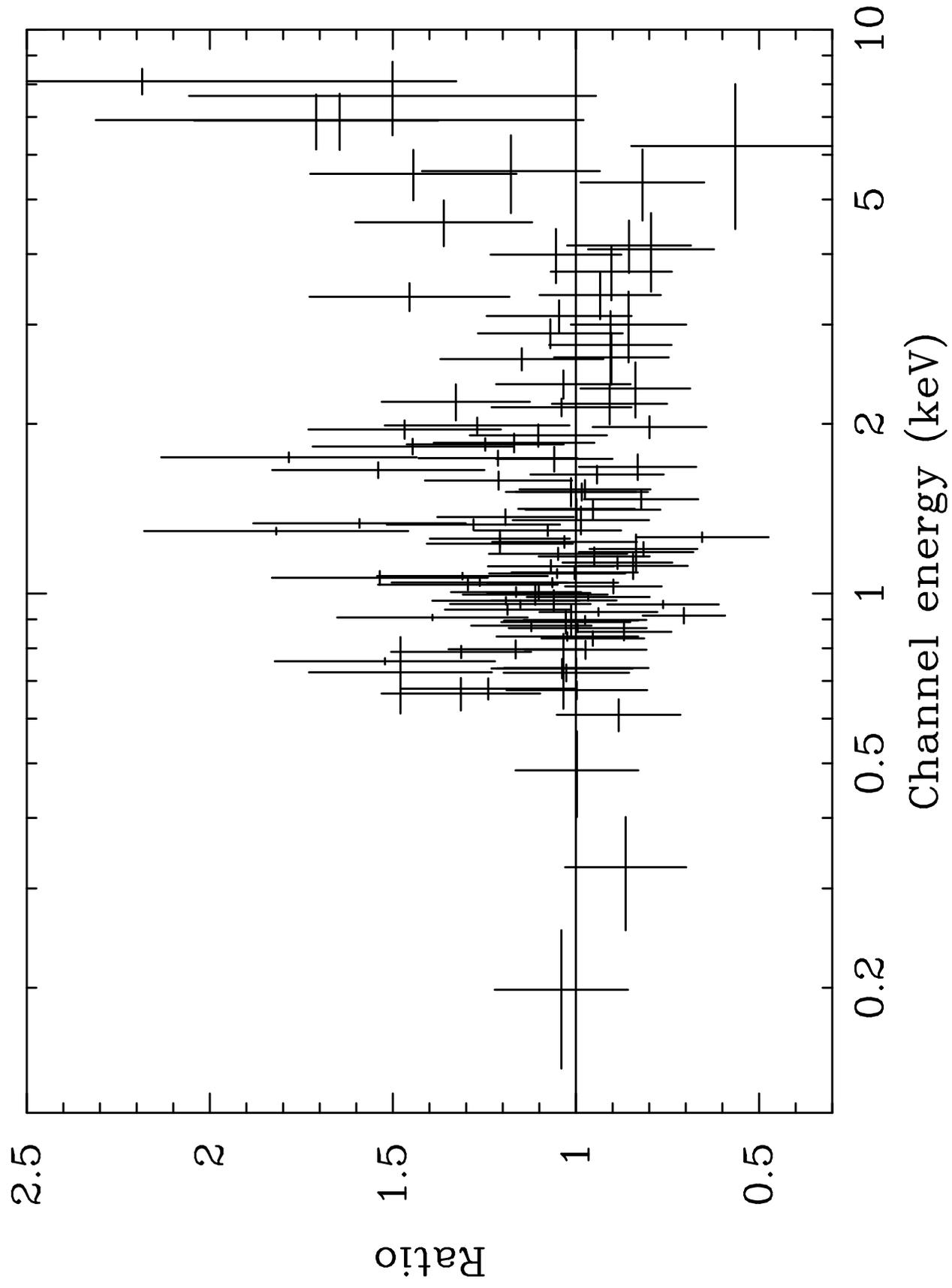
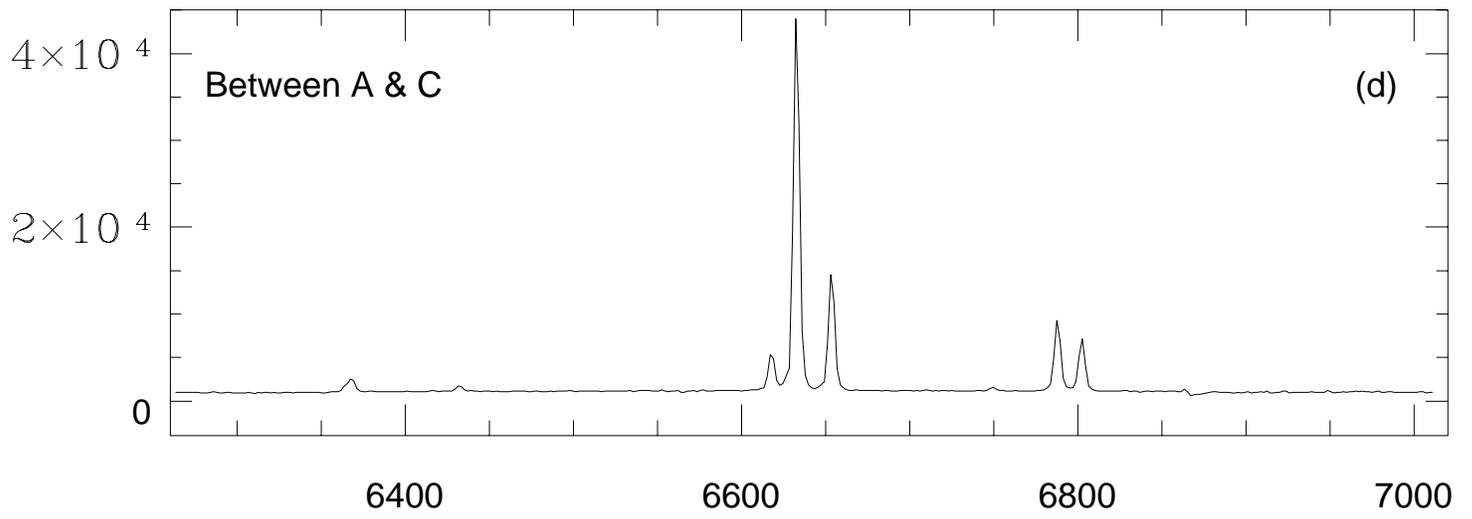
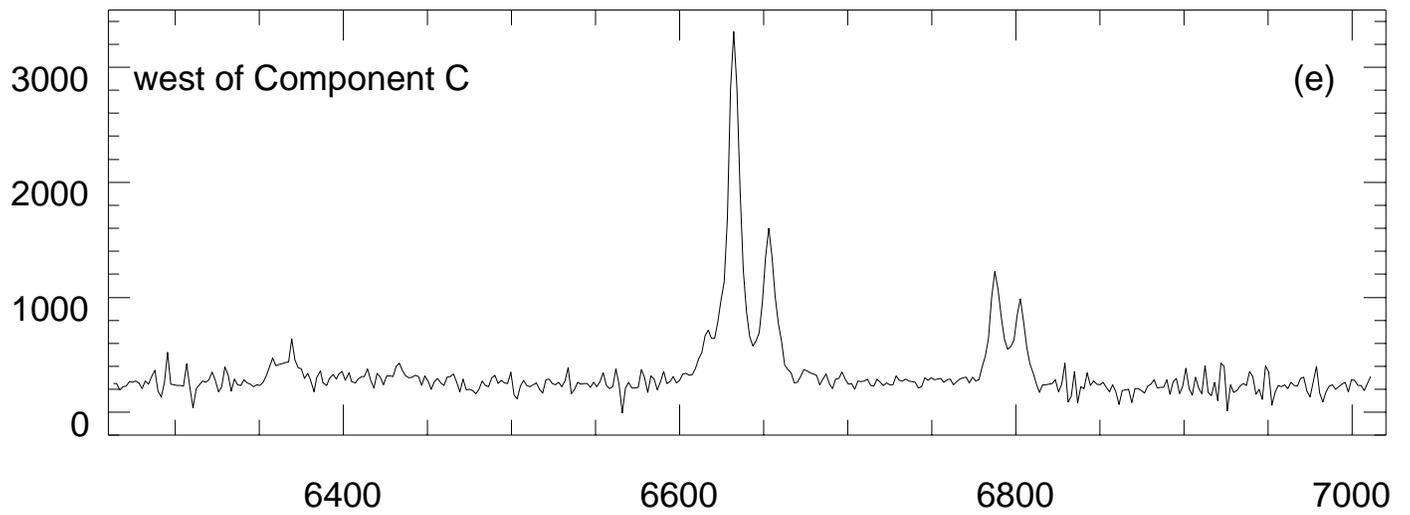
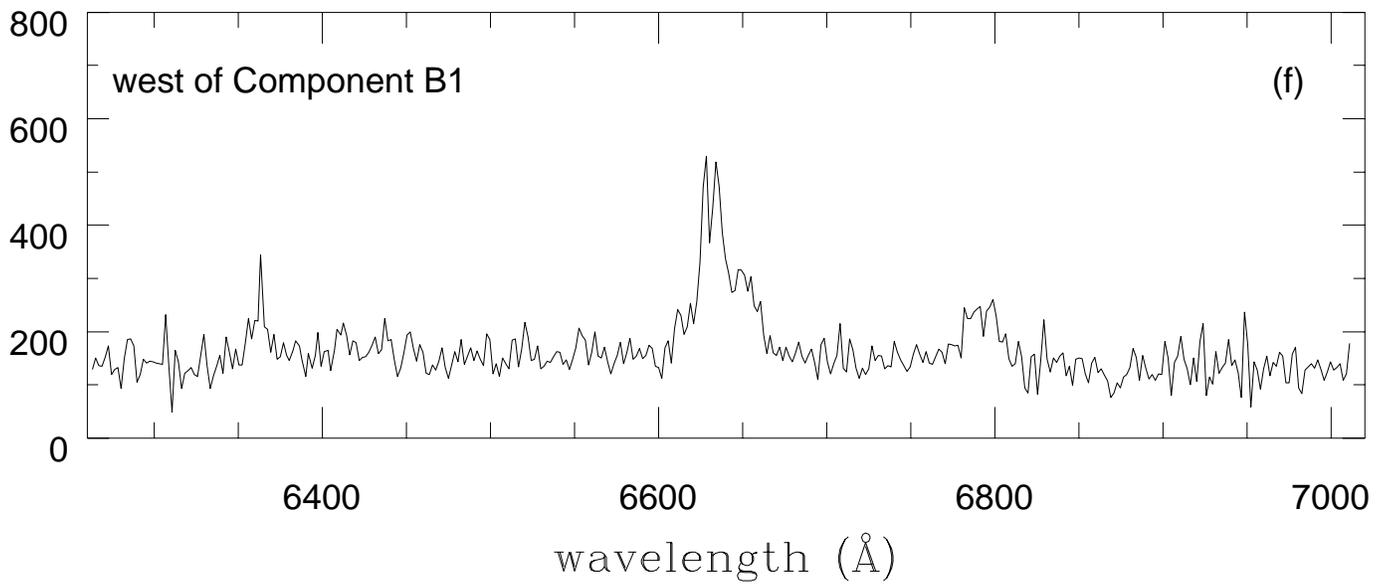
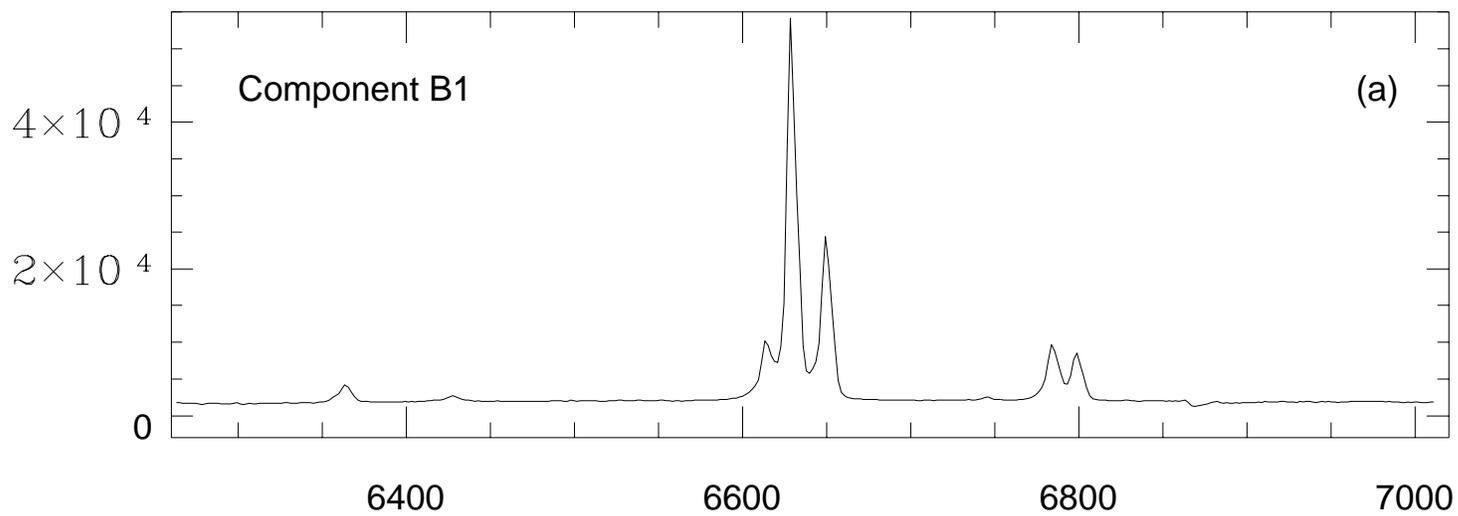
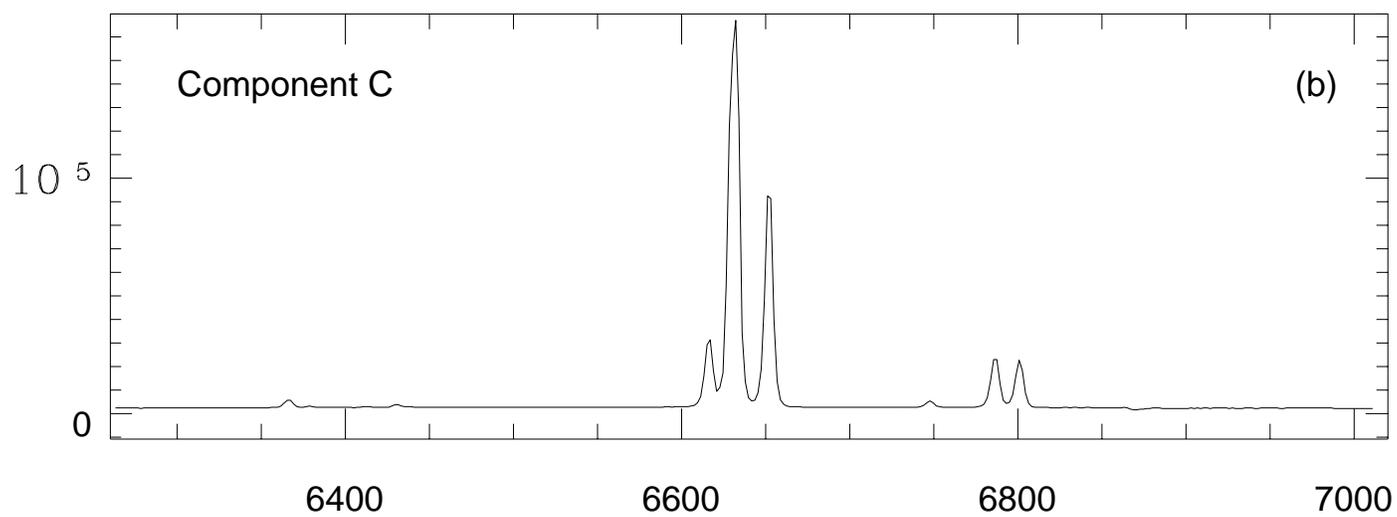
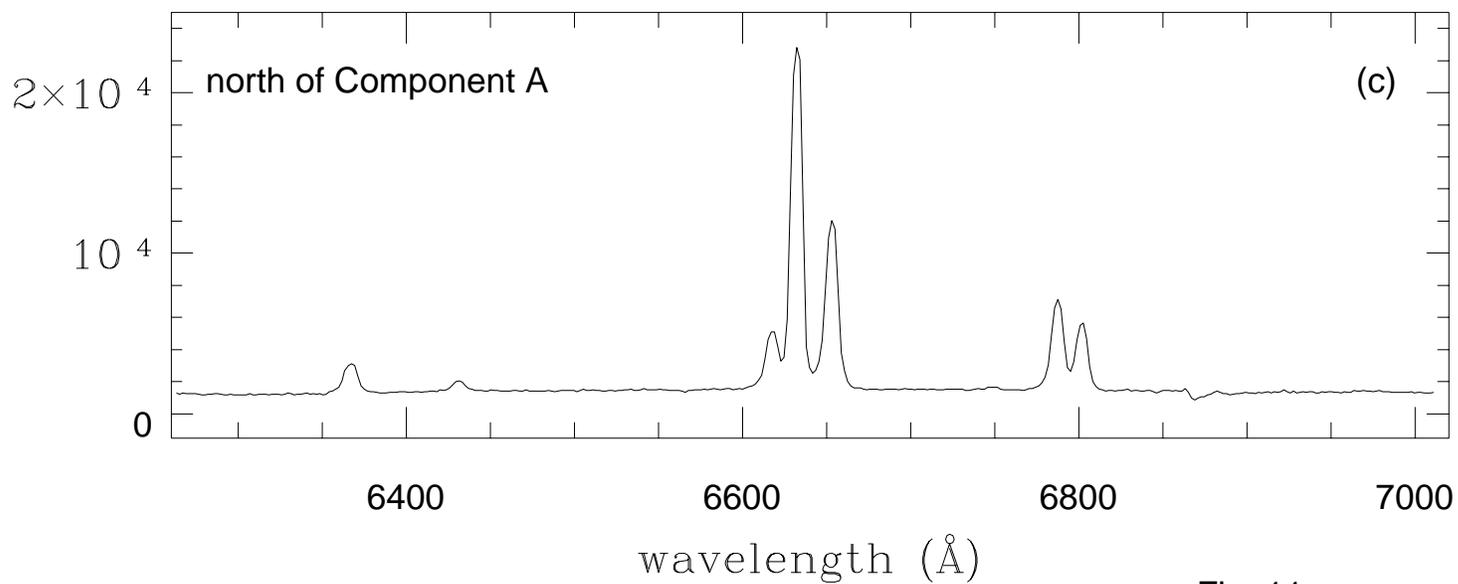
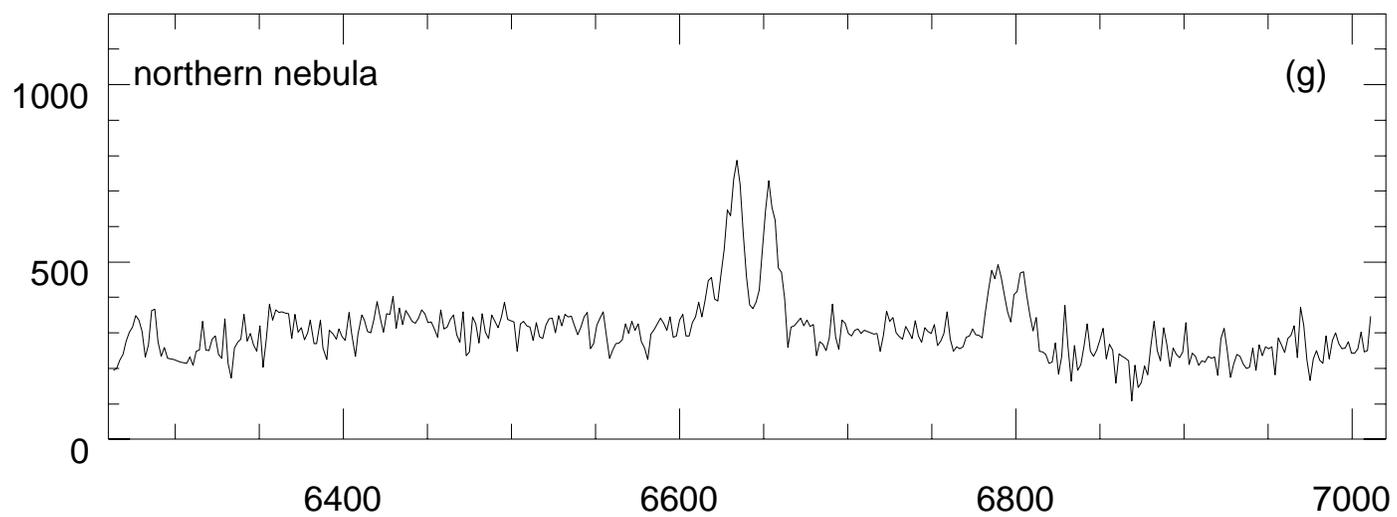
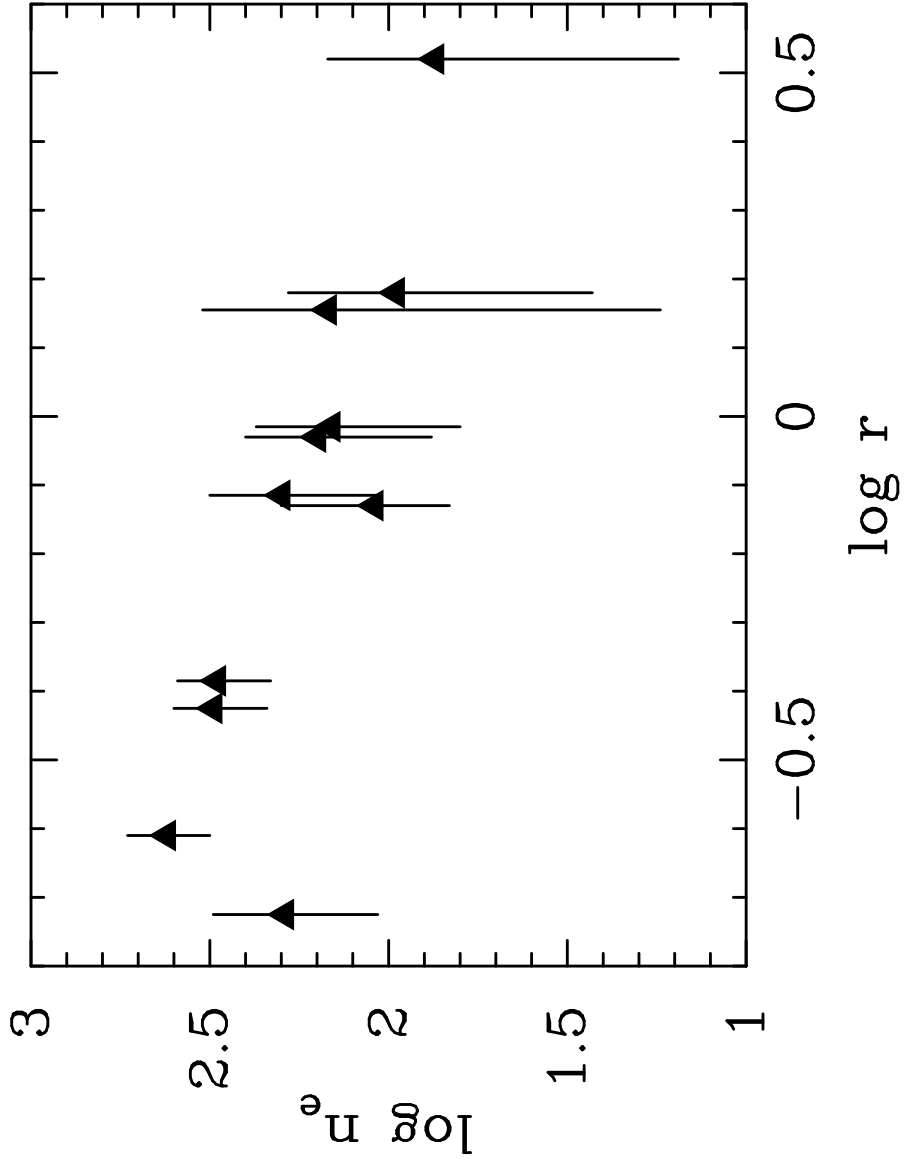
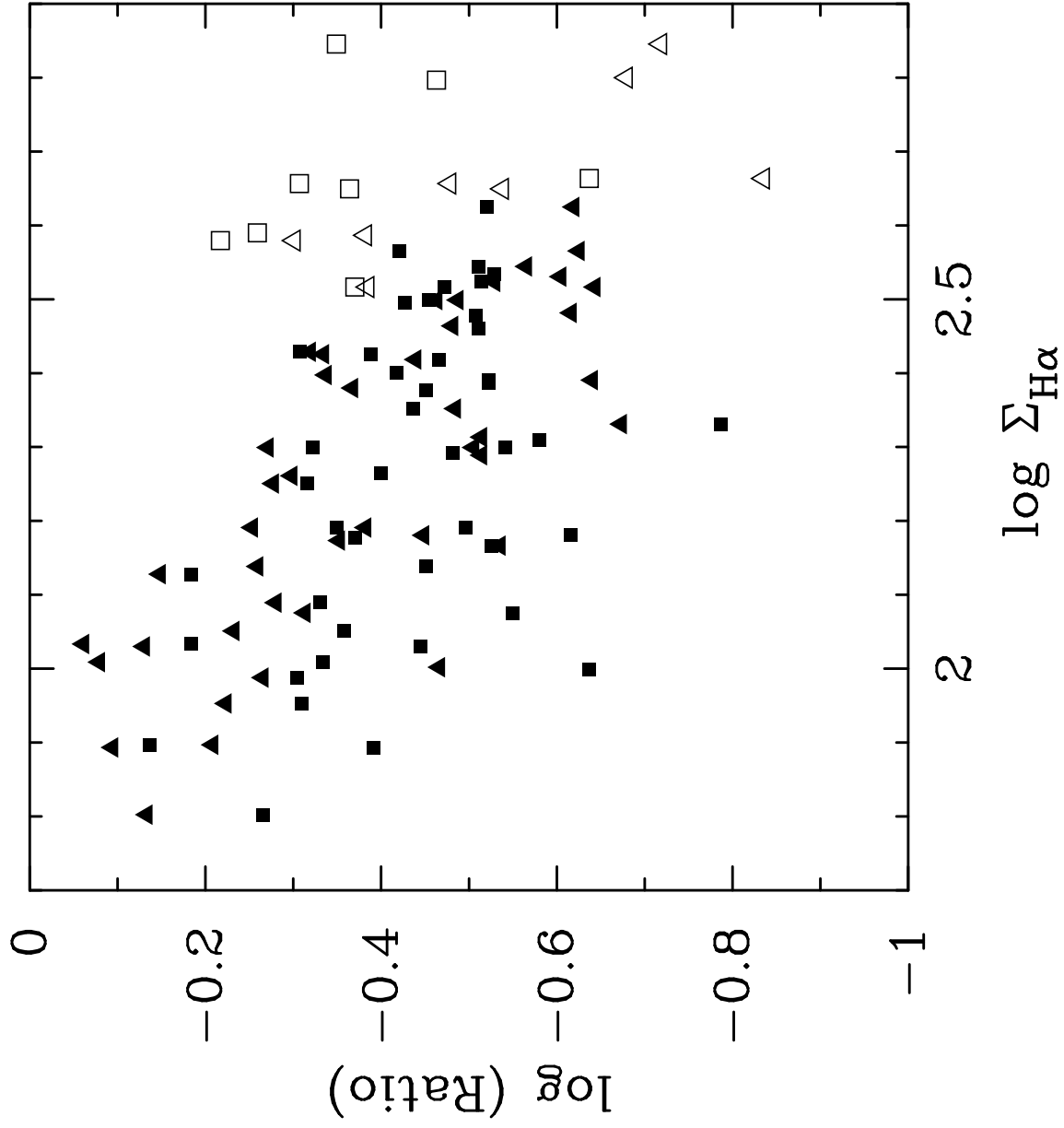
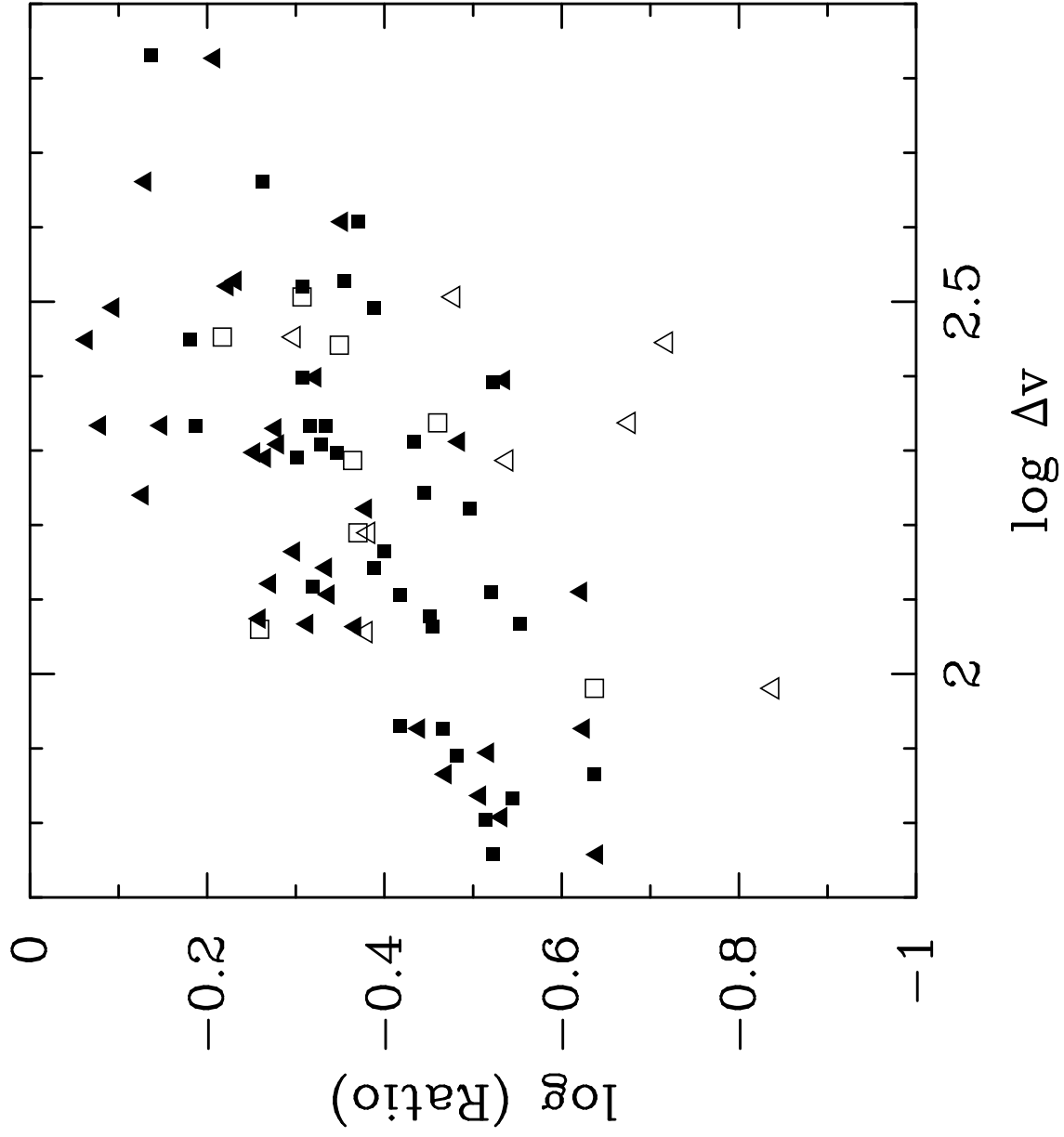
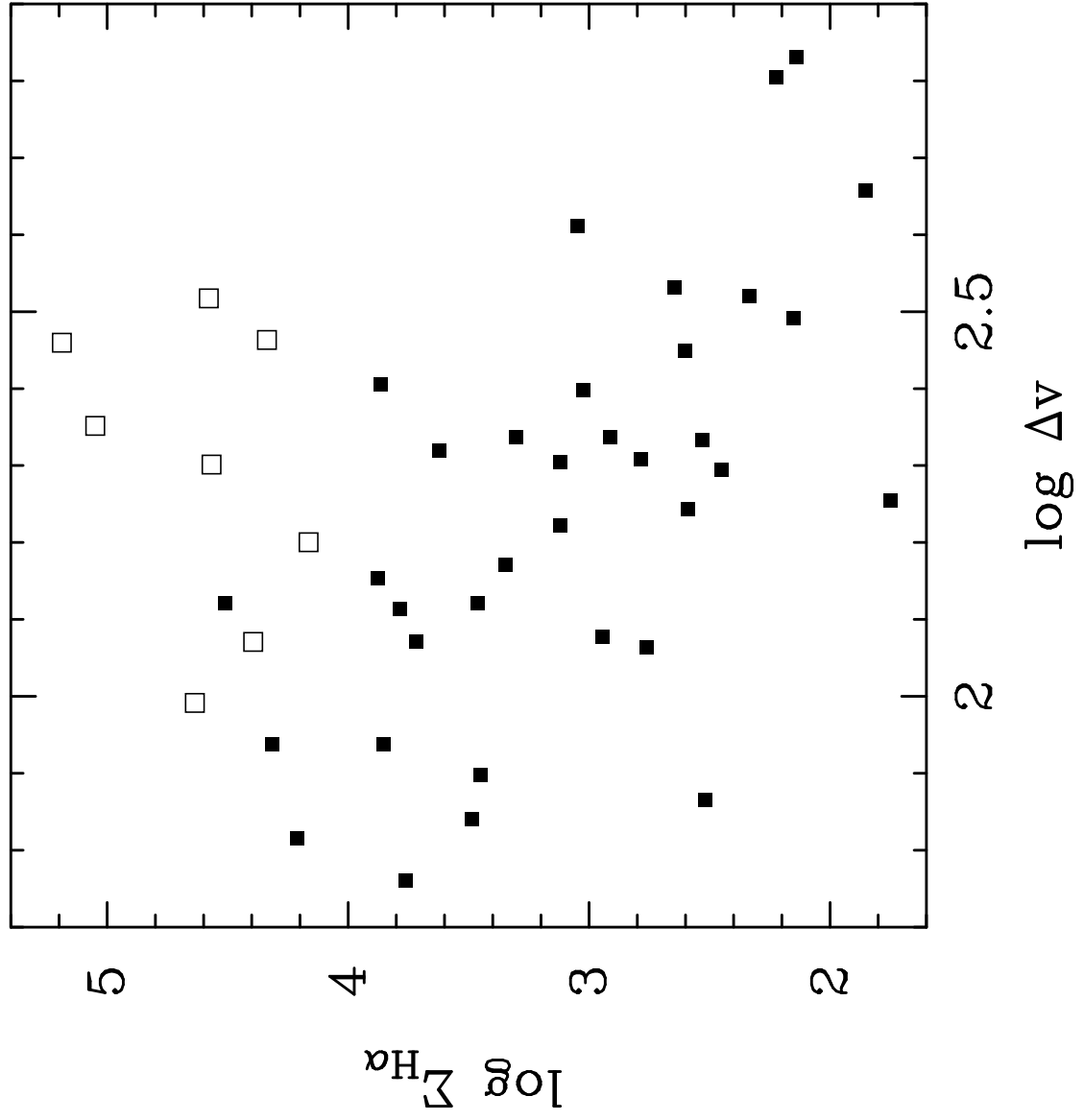
\end{document}